\documentclass[preprint,aps,amsmath,amssymb,superscriptaddress,floatfix]{revtex4-2}

\usepackage[utf8]{inputenc}
\usepackage{setspace}
\usepackage{graphicx}
\usepackage{color}
\usepackage{soul}
\usepackage{hyperref}
\usepackage{miller}
\usepackage{pdfpages}
\usepackage{bbold}
\usepackage{gensymb}
\usepackage{xr}

\newcommand{\simlt}
      {\ifmmode       { \raisebox{-.8em}{$<$}\atop\sim}
         \else        {$\raisebox{-.8em}{$<$}\atop\sim$}
      \fi}

\makeatletter
\def\@hangfrom@section#1#2#3{\@hangfrom{#1#2}#3}
\def\@hangfroms@section#1#2{#1#2}
\makeatother

\makeatletter
\AtBeginDocument{\let\LS@rot\@undefined}
\makeatother

\usepackage[version=4]{mhchem}

\begin{document}


\title{Imaging stripe dynamics in trilayer nickelate \ce{La4Ni3O10}}

\author{Uladzislau Mikhailau}
\affiliation{SUPA, School of Physics and Astronomy, University of St Andrews, North Haugh, St Andrews, KY16 9SS, United Kingdom}

\author{Luke C. Rhodes}
\affiliation{SUPA, School of Physics and Astronomy, University of St Andrews, North Haugh, St Andrews, KY16 9SS, United Kingdom}

\author{Siri A. Berge}
\affiliation{SUPA, School of Physics and Astronomy, University of St Andrews, North Haugh, St Andrews, KY16 9SS, United Kingdom}

\author{Matthias Hepting}
\affiliation{Max-Planck-Institute for Solid State Research, Heisenbergstr. 1, 70569 Stuttgart, Germany}

\author{Masahiko Isobe}
\affiliation{Max-Planck-Institute for Solid State Research, Heisenbergstr. 1, 70569 Stuttgart, Germany}

\author{Carolina de Almeida Marques}
\affiliation{SUPA, School of Physics and Astronomy, University of St Andrews, North Haugh, St Andrews, KY16 9SS, United Kingdom}

\author{Pascal Puphal}
\affiliation{Max-Planck-Institute for Solid State Research, Heisenbergstr. 1, 70569 Stuttgart, Germany}

\author{Peter Wahl}
\email[Correspondence to: ]{wahl@st-andrews.ac.uk.}
\affiliation{SUPA, School of Physics and Astronomy, University of St Andrews, North Haugh, St Andrews, KY16 9SS, United Kingdom}
\affiliation{Physikalisches Institut, Universität Bonn, Nussallee 12, 53115 Bonn, Germany}

\date{\today}

\begin{abstract}
Since the discovery of high-temperature superconductivity in nickelate superconductors, it is an open question how closely the superconducting state resembles that of cuprate superconductors. One salient feature of the phase diagram of the high-temperature cuprate superconductors is stripe order. Despite their prevalence, real-space imaging has been limited to the charge sector. Here we use spin-polarised scanning tunnelling microscopy to visualize the local magnetic and charge distribution emerging due to a stripe order in the trilayer nickelate \ce{La4Ni3O10}. The stripe order exhibits a four unit cell periodicity, closely resembling that seen in cuprates, and opens a near-complete $\sim66\mathrm{meV}$ gap at the Fermi level. Crucially, discrete phase slips can be triggered by tunneling electrons above a $\sim 20\mathrm{meV}$ threshold, allowing imaging of stripe dynamics at the atomic scale. These results highlight the importance of correlation physics driving stripe-like orders in lanthanum nickelates with striking similarities to the cuprates.
\end{abstract}


\maketitle

\section{Introduction}
\begin{figure}
\includegraphics[width=\textwidth]{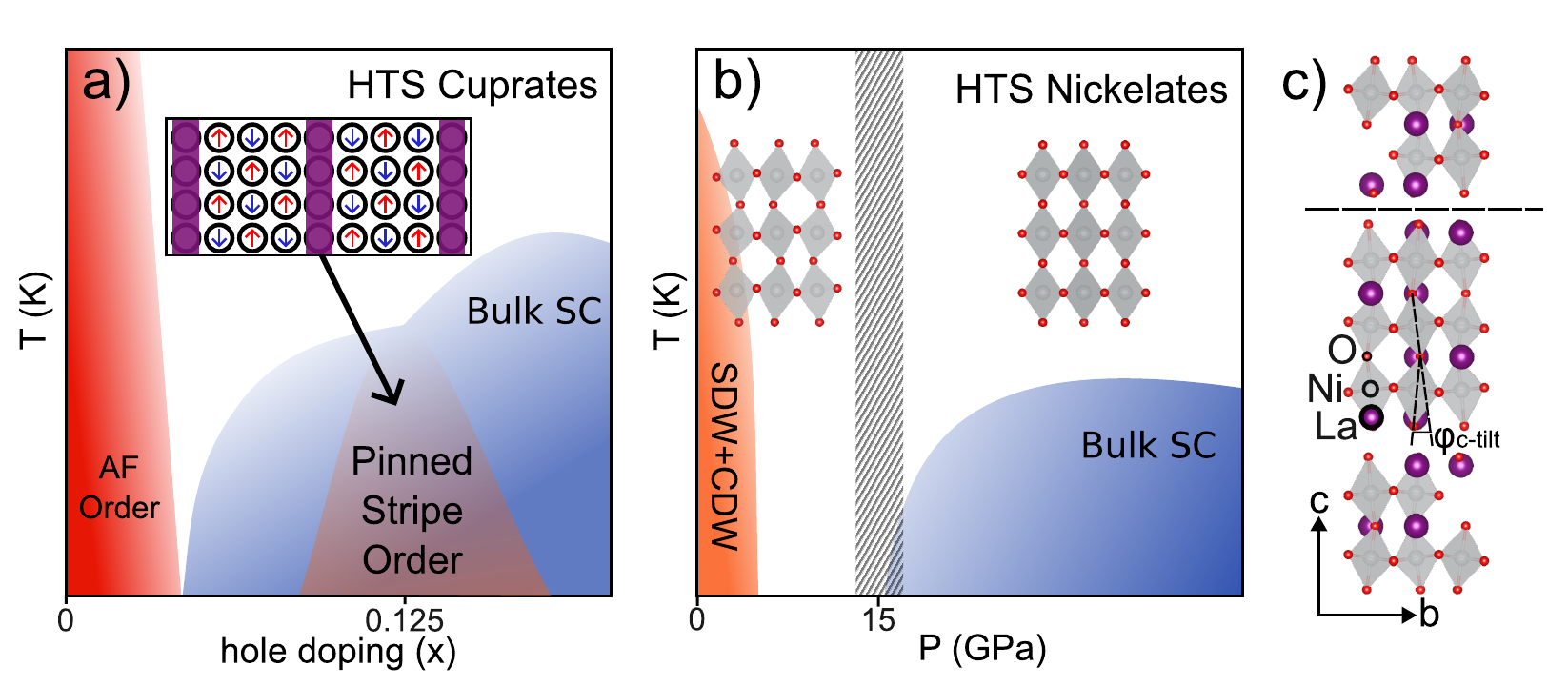}
    \caption{\textbf{Stripe orders in Mott insulators.} (a) Generic phase diagram of the cuprate superconductors. Stripe fluctuations freeze into static stripe order close to $x=1/8$. Inset shows a sketch of pinned stripe order, consisting of antiphase antiferromagnetic domains separated by lines of mobile charge. (b) Schematic phase diagram of \ce{La4Ni3O10} with pressure. Under ambient conditions, the compound behaves as a normal metal with a spin-density-wave-like transition at low temperature. Application of pressure triggers a structural phase transition which straightens out the Ni-O-Ni bonds, similar to \ce{La3Ni2O7}, and high-temperature superconductivity is observed. (c) Crystal structure of \ce{La4Ni3O10} in the low-pressure low-symmetry phase, with the preferred cleavage plane indicated by a dashed line (La - purple, Ni - grey, O - red) and crystallographic direction defined as used throughout this work.}
    \label{structure4310}
\end{figure}

It has long been a riddle why high temperature superconductivity only occurs in cuprate materials. The recent discovery of superconductivity in \ce{La3Ni2O7} \cite{sun_signatures_2023} has introduced a new class of high-temperature superconductors with very similar structure but different electronic configuration. One of the most pressing questions is whether the two classes of materials share a similar pairing mechanism and correlated electron physics. Identifying the commonalities promises new insights into the mechanisms of high-temperature superconductivity in both.
Many aspects of the phase diagram of high temperature cuprate superconductors (Fig.~\ref{structure4310}a) can be captured based on a Mott-Hubbard Hamiltonian, where correlated phases emerge close to half filling as a consequence of the interplay between localization due to on-site Coulomb repulsion and itineracy due to doping.
Early on it was found that the Mott-Hubbard Hamiltonian exhibits an instability towards stripe order accompanied by spin-charge separation\cite{zaanen_charged_1989,kivelson_electronic_1998}. Subsequently, such stripe orders have been observed experimentally in \ce{La2NiO4}\cite{tranquada_cooperative_1995} as well as in the high-tc cuprate superconductors \cite{tranquada_coexistence_1997}. 
In some cases, these stripe orders are fluctuating and coexist with superconductivity\cite{tranquada_coexistence_1997,tranquada_evidence_2008,ghiringhelli_long-range_2012}. Multiple theories argue that stripe orders play an important role in shaping the superconducting phase diagram of the cuprate high-temperature superconductors.\cite{emery_importance_1995,loder_modeling_2011,jiang_stripe_2022} At specific dopings, the stripe order becomes commensurate with the lattice and competes with the superconducting state. Such a stripe order is characterized by antiferromagnetic domains separated by one dimensional lines of doped charge (see sketch in Fig.~\ref{structure4310}a). The direction of the stripes is primarily defined by the symmetry of the crystal structure, as observed experimentally in doped perovskite cuprates and nickelates\cite{hucker_coupling_2007}. 

In cuprate high temperature superconductors, the stripe orders are typically short-ranged and locally commensurate. Scanning tunneling microscopy has been a key tool to characterize this short-ranged stripe order\cite{hanaguri_checkerboard_2004,kohsaka_intrinsic_2007}. Charge modulations with a similar periodicity have now been observed across a number of cuprate superconductors at different dopings\cite{ghiringhelli_long-range_2012,comin_charge_2014}, suggesting that they are a common phenomenon in these materials.

In the nickelates, both the bi- and trilayer compounds \ce{La3Ni2O7} and \ce{La4Ni3O10} (Fig.~\ref{structure4310}b) become superconducting under high pressure.\cite{sun_signatures_2023,wang_pressure-induced_2024,zhu_superconductivity_2024} Similar to \ce{La3Ni2O7}, at ambient pressure \ce{La4Ni3O10} is reported to host a spin-density wave phase\cite{zhang_intertwined_2020} accompanied by a structural distortion associated with octahedral tilts. When subjected to hydrostatic pressure, these tilts straighten out (Fig.~\ref{structure4310}(c)) as a structural phase transition occurs before entering a tetragonal phase that exhibits bulk superconductivity\cite{zhu_superconductivity_2024}. Transport and ARPES experiments on nickelates show signatures of Mott-Hubbard-physics\cite{kobayashi_transport_1996} with strongly orbital dependent correlations\cite{yang_orbital-dependent_2024,li_orbital-selective_nodate,sohn_layer-controlled_2025}, although here in a multi-band setting.

Here, we image stripe order and dynamics in \ce{La4Ni3O10} using low temperature scanning tunneling microscopy and spectroscopy. Our data show complex charge modulations associated with this stripe order, with energy-dependent localization of the charge density and formation of a gap that almost completely depletes the differential conductance close to the Fermi energy. We use spin-polarized STM to demonstrate that these stripes consist of intertwined magnetic and charge orders. Our results suggest that the stripe-like order emerges due to a complex interplay of nesting instability, magnetic order, electronic correlations and lattice distortions.

\section{Results}
The crystal structure of \ce{La4Ni3O10} has a natural cleavage plane between \ce{LaO} planes (dashed line in Fig.~\ref{structure4310}(c)) which is expected to result in similarly high-quality surfaces as seen in the cuprates\cite{hudson_interplay_2001} or  ruthenates\cite{marques_magneticfield_2021}. Following sample cleavage, we observe only one type of termination with consistent step height, work function and differential conductance spectrum (see Suppl. Fig.~S1).

\begin{figure}
\includegraphics[width=\textwidth]{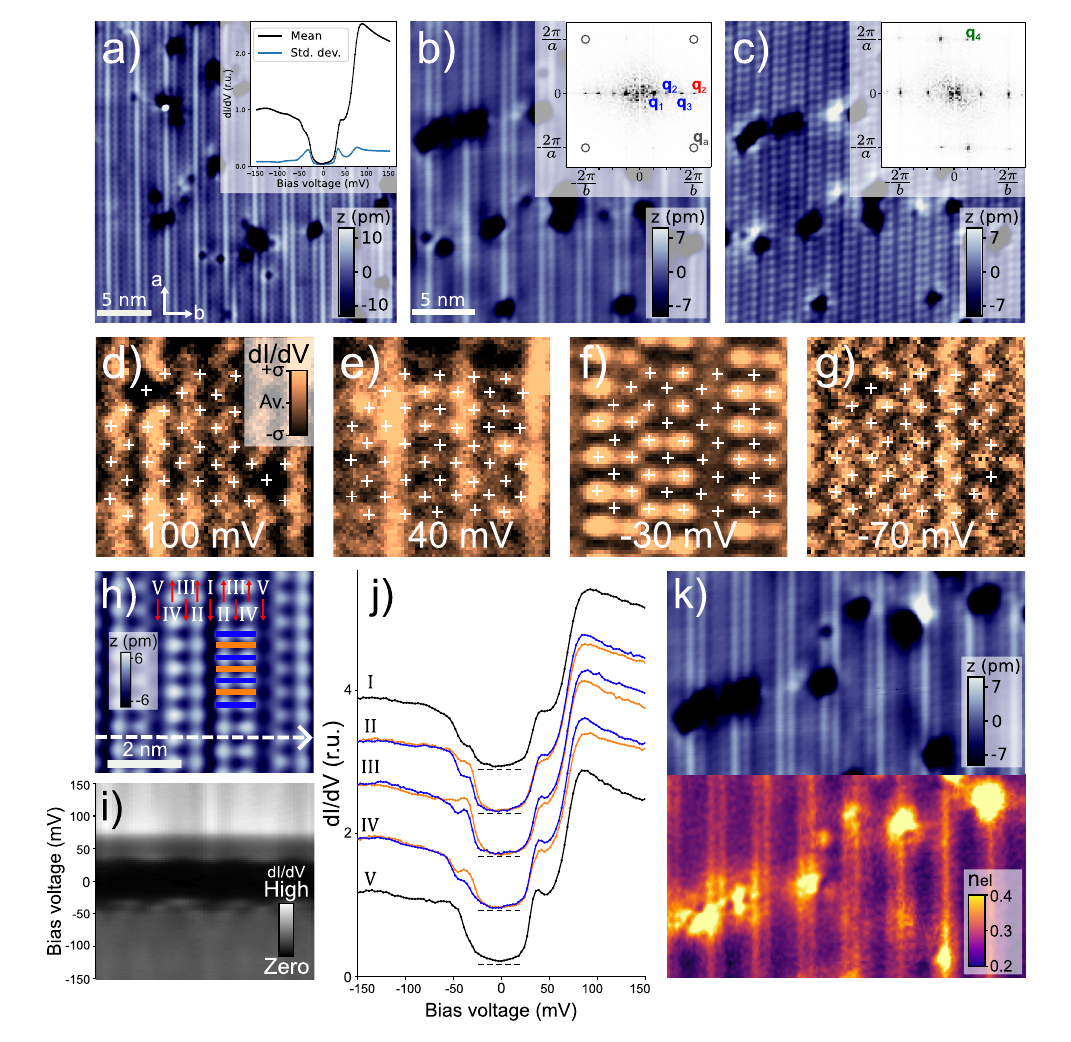}
    \caption{\textbf{Charge stripe order in \ce{La4Ni3O10.}} (a) Large scale topography showing locally commensurate stripe order ($V_\mathrm{s}=150~\mathrm{mV}$, $I_\mathrm{s} = 30~\mathrm{pA}$). Inset shows a spatially averaged differential conductance spectrum $g(V)$ and its standard deviation over multiple unit cells of the stripe order. (b, c) Large scale topographies at positive and negative bias voltage and corresponding FFTs in the insets ($V_\mathrm{s} = 150, -150~\mathrm{mV}$, $I_\mathrm{s}=40~\mathrm{pA}$, taken in the same area with the same tip). (d-g) Constant height differential conductance maps $g(V,\mathbf{r})$ of the charge order. Crosses indicate positions of nickel atoms. (h) Topography from a different area ($V_\mathrm{s} = -150\mathrm{mV}$, $I_\mathrm{s} = 30~\mathrm{pA}$) where detailed tunneling spectra have been recorded. (i) Line of $g(V,r)$ spectra acquired along the dashed line in (h) (lateral dimension same as in (h), $V_\mathrm{s} = -150\mathrm{mV}, I_\mathrm{s} = 40\mathrm{pA}$). (j) $g(V,\mathbf{r})$ spectra averaged over symmetry equivalent lines indicated by roman numerals and red arrows in (h). Blue and orange colored spectra are recorded at positions of blue and orange bars in h where spectra exhibit differences along the vertical direction in (h) ($V_\mathrm{s} = -150~ \mathrm{mV}$, $I_\mathrm{s} = 40\mathrm{pA}$).
    (k) Topography $z(\mathbf{r})$ aligned with a map of the doped electron density $n_\mathrm{el}(\mathbf{r})$ at $V=81~\mathrm{mV}$ (see suppl. section S2 for details). Lines of increased electron density are out of phase with the charge stripe modulation visible in the topography. }
    \label{chargemodulation}
\end{figure}

STM images exhibit a clear stripe-like charge modulation (see Fig. \ref{chargemodulation} (a-c)) while tunneling spectra exhibit a gap around the Fermi energy (see inset in Fig.\ref{chargemodulation}(a)), consistent with previous observations\cite{li_direct_2025}. The size of the gap is about $66\mathrm{meV}$, with a small residual density of states within the gap. The spectra have a number of distinct peaks, particularly a pair of peaks directly around the gap. The intensity of these peaks varies significantly depending on the position within the stripe unit cell, as seen in the standard deviation of the differential conductance spectrum (inset in Fig.\ref{chargemodulation}(a)). The octahedral tilting in \ce{La4Ni3O10} leaves clear traces in STM images as a "zigzag" modulation, similar to what is observed in ruthenates\cite{halwidl_ordered_2017}, visible in the Fourier transformation as peaks at $\mathbf{q}_z=(0,1)$ between the atomic ones at $\mathbf{q}_\mathrm{a}=(1,1)$ and $(1,-1)$, Fig.~\ref{chargemodulation}(b-c)).

At positive bias voltages, Fig.~\ref{chargemodulation}(b), the stripe modulation has a characteristic appearance with a four-unit cell periodicity. This is reflected in a dominant peak at $\mathbf{q}_1=(0,1/4)$ in the corresponding FFT image (Fig.~\ref{chargemodulation}b). Rather than being represented by a single peak in the FFT, the modulation of the charge density consists of a set of periodicities that are multiples of $\mathbf{q}_1=(0, 1/4)$ at $\mathbf{q}_2=(0,2/4)$ and $\mathbf{q}_3=(0,3/4)$, originating from a quasi-square-shaped modulation of the charge density. At negative bias voltages, an additional dominant modulation at $\mathbf{q}_4=(1,1/4)$ appears (see Fig. \ref{chargemodulation}(c)). 
The order is locally commensurate, in Suppl. Fig.~S2 we highlight regions with different local commensuration.

The complex energy-dependence of the stripe contrast is also seen in differential conductance maps $g(\mathbf{r},V)$: Figs.~\ref{chargemodulation}(d-g) show a set of constant-height differential conductance $g(\mathbf{r},V)$ maps recorded at different bias voltages. The images show significant changes within a small energy range. Fig.~\ref{chargemodulation}(d), taken at a high positive bias voltage $V$ above the gap, shows the zigzag modulation due to the octahedral tilts. Comparison with calculations (see Suppl. Fig.~S3) allows determining the positions of the nickel atoms (white crosses). The charge stripe modulation becomes more visible close to the gap edge at $V= 40 ~\mathrm{mV}$, Fig.~\ref{chargemodulation}(e). It is apparent that the stripes are centered between the nickel atoms, on top of in-plane oxygen atoms. Between unoccupied and occupied states, we observe a dramatic change in contrast and a change to a dominant $(1/4,1)$ modulation, Fig.~\ref{chargemodulation}(f), before reverting to the zigzag modulation (Fig.~\ref{chargemodulation}(g)).
The complexity of the spatial patterns with bias voltage is also reflected in tunneling spectra: the spectra show characteristic changes across the stripes. In particular, the shoulders at around $\pm40\mathrm{meV}$ change significantly while moving across the stripe order. Figures~\ref{chargemodulation}(h-j) show averaged differential conductance spectra along symmetry-equivalent vertical lines on top of two stripes. While in some regions, the spectra hardly change along the stripes (black spectra in Fig.~\ref{chargemodulation}(j)), in other areas there is a significant modulation as shown by the blue and orange spectra. The peak at $-33\mathrm{mV}$ shifts between $-30\mathrm{mV}$ and $-36\mathrm{mV}$. The peak at positive bias voltages changes significantly less both across and along the stripes. 

In the Mott-Hubbard picture, the marked particle-hole asymmetry of the spectra can be interpreted as a consequence of the doping and the corresponding fugacity $Z(V)$ (see suppl. sect.~S2).\cite{anderson_theory_2006} Here, the known fractional valence of Ni in \ce{La4Ni3O10} implies electron doping of the system\cite{periyasamy_effect_2021}, hence a higher probability for electron extraction, i.e. higher differential conductance at positive bias voltages is expected. In Fig.~\ref{chargemodulation}(k), we show a spatial map of 
the electron doping $n_\mathrm{el}$ extracted from the fugacity (See Suppl. Sect.~S2 for definition) showing a local variation consistent with increased doping between the maxima of the stripes. 

\begin{figure}
\includegraphics[width=0.5\textwidth]{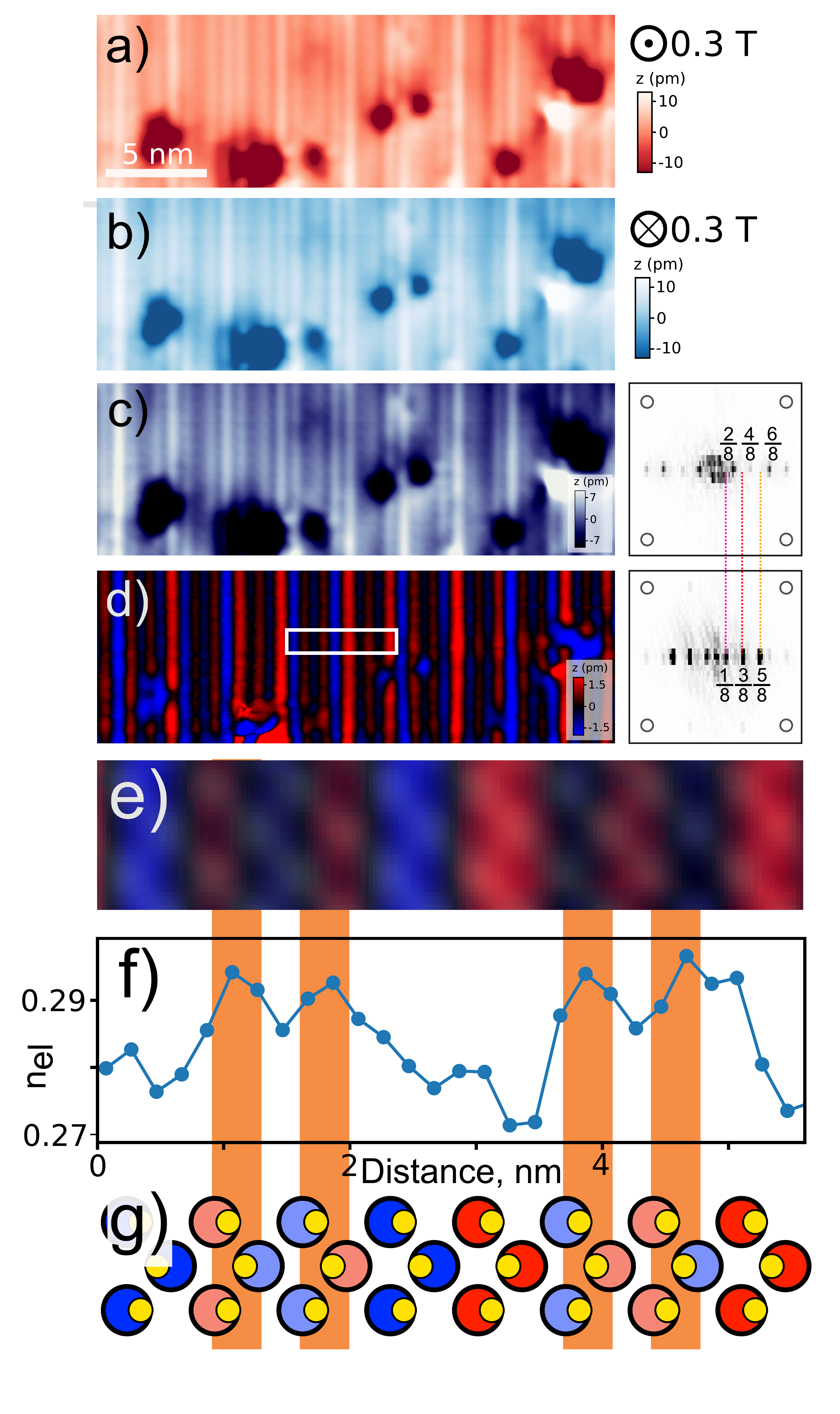}
    \caption{\textbf{Magnetic stripe order in \ce{La4Ni3O10}.} (a, b) Topographic images $z(\mathbf{r},H)$ recorded with a spin-polarized STM tip in a magnetic field of $\mu_0H=+/-0.3\mathrm{T}$. A change in the stripe contrast can be clearly seen ($V_\mathrm{s} = 150\mathrm{mV}, I_\mathrm{s} = 30\mathrm{pA}$). Images in (c, d) show the sum $\Sigma_z(\mathbf{r},H)=(z(\mathbf{r},H)+z(\mathbf{r},-H))/2$ (c)  showing only the charge order and the difference $\Delta_z(\mathbf{r},H)=(z(\mathbf{r},H)-z(\mathbf{r},-H))/2$ (d) exhibiting purely the magnetic contrast. Next to (c, d), we show the Fourier transformation, revealing a signal at $q=2/8$, $4/8$ and $6/8$ for the charge contrast and at $q=1/8$, $3/8$ and $5/8$ for the magnetic contrast. (e) shows the magnetic contrast superimposed on a Fourier filtered image of the nickel lattice for the region indicated by the white box in (d). (f) shows the electron density $n_\mathrm{el}(V=90\mathrm{mV})$ estimated from the ratio of the differential conductances $g(-V,
    \mathbf{r})/g(V,\mathbf{r})$ (see suppl. sect.~S2) across the same region as shown in (e), vertical orange bars highlighting positions of electron-doped lines. (g) shows a model extracted from panels (a-f) and Fig.~\ref{chargemodulation}, superimposing the doping (orange lines) and magnetic order on the nickel sites (red/blue colored circles). Yellow circles indicate apical oxygen. }
    \label{magneticorder}
\end{figure}

One of the most important consequences of the spin-charge separation of the stripe orders in cuprates is that there are magnetically ordered regions interdigitated with the charged stripes. Such magnetic order can be detected in STM using a magnetic tip\cite{wiesendanger_spin_2009,enayat_real-space_2014}. Fig.~\ref{magneticorder}(a, b) show images recorded with a ferromagnetic tip in opposite magnetic fields, which switches the magnetization of the tip but not that of the sample. It reveals significant changes reflecting the magnetic order accompanying the stripe order. This can be more clearly seen by looking at the sum $\Sigma$ and the difference $\Delta$ of the two images, Fig.~\ref{magneticorder}(c, d), which reflect the charge (c) and spin (d) contrast. The charge contrast looks identical to that seen with a non-magnetic tip, compare Fig.~\ref{chargemodulation}(b). The magnetic contrast, (d), shows regions of strong magnetic contrast alternating with regions of weak magnetic contrast. The ordering vectors of the magnetic order are $1/8$, $3/8$ and $5/8$ (compare Fourier transformation in (d)). The charge order, (inset in (c)), shows twice the wave vector of the magnetic order. 
In Fig.~\ref{magneticorder}(e), we superimpose the magnetic order seen in spin-polarized images with a Fourier-filtered topography showing the nickel lattice. 
It can be seen that the magnetic order consists of double-stripe domains with only small admixture of not fully developed single stripe regions. 
Regions of low magnetic contrast appear to be electron-rich as deduced from the local doping $n_\mathrm{el}$ extracted from the ratio $Z^\prime(V,\mathbf{r})=g(-V,
    \mathbf{r})/g(V,\mathbf{r}$ (see suppl. sect.~S2)  (Fig.~\ref{magneticorder}(f)).
Fig.~\ref{magneticorder}(g) shows a model of this double-stripe order summarizing the experimental findings: We find antiphase antiferromagnetic domains separated by quasi-1D electron-rich rivers, where magnetic contrast is significantly suppressed.

\begin{figure}
\includegraphics[width=\textwidth]{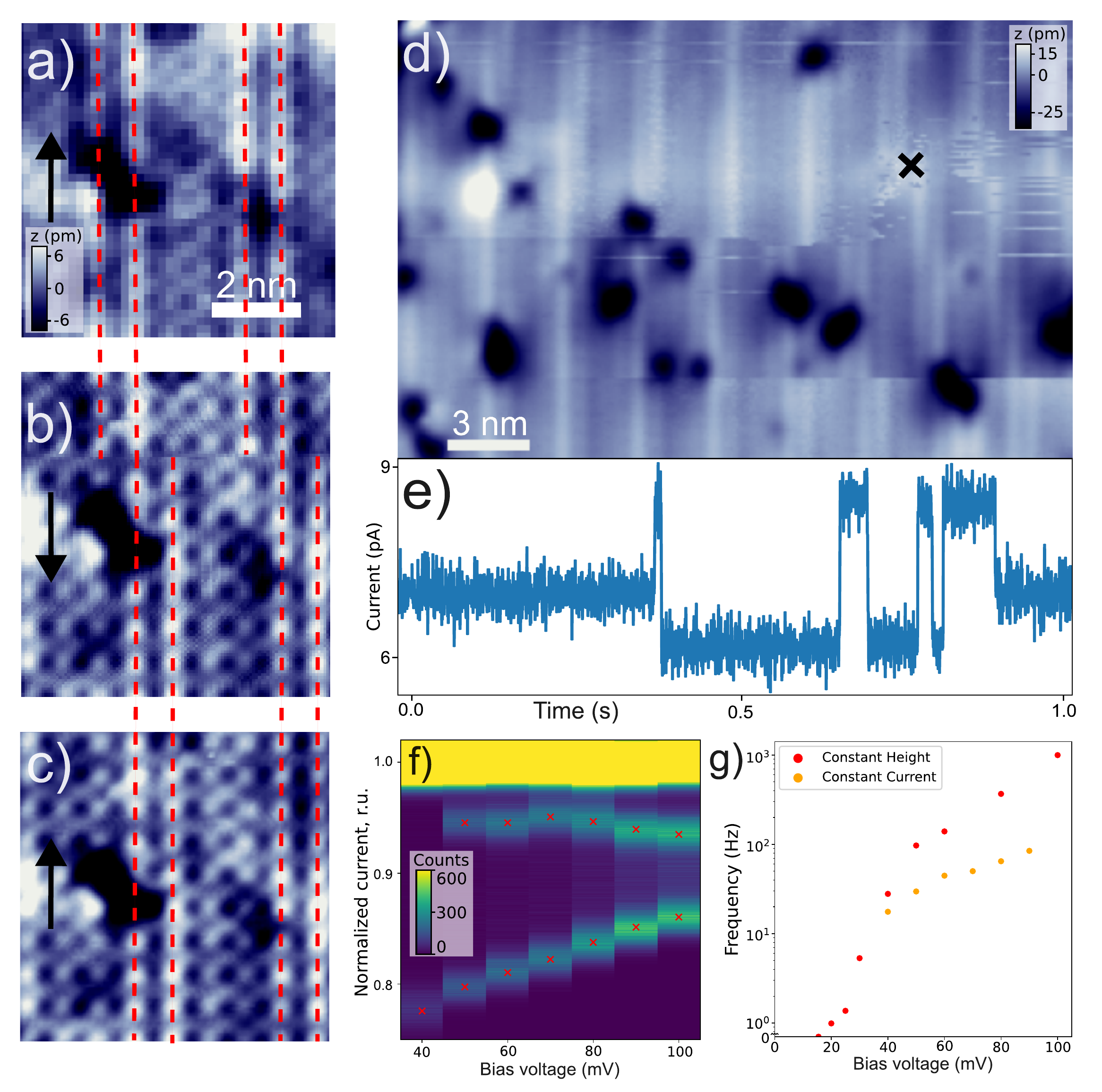}
    \caption{\textbf{Stripe Dynamics} (a-c) Small scale STM topographies taken with a non-magnetic tip before (a), when (b), and after (c) a stripe moved ($V_\mathrm{s} = 200 \mathrm{mV}$, $I_\mathrm{s} = 10 \mathrm{pA}$). (d) Large scale STM topography  showing tip induced stripe fluctuations ($V_\mathrm{s} = 100 \mathrm{mV}$, $I_\mathrm{s} = 30~\mathrm{pA}$). The cross depicts the position where the time resolved data was collected. (e) Time trace of the current showing telegraph noise originating from stripe fluctuations ($V=20\mathrm{mV}$). (f) Histogram of the current trace for different bias voltages $V$ showing three fluctuating states taken with constant current ($I_\mathrm{s} = 30~\mathrm{pA}$). (g) Dependence of the switching frequency on the bias voltage $V$ in constant height mode after stabilizing the tip-sample distance at $V_\mathrm{s} = 40~\mathrm{mV}$ and $I_\mathrm{s} = 30~\mathrm{pA}$, and in constant current mode for $I_\mathrm{s} = 30~\mathrm{pA}$. }
    \label{stripedynamics}
\end{figure}

We occasionally observe spontaneous phase jumps of the stripe order by a unit cell, see Fig.~\ref{stripedynamics}(a-d). Fig.~\ref{stripedynamics}(a) shows an image recorded before a phase jump has occurred, in Fig.~\ref{stripedynamics}(b) a jump occurred while the image was being recorded. In images taken afterwards (Fig.~\ref{stripedynamics}(c)), the whole stripe order has changed its phase. Fig.~\ref{stripedynamics}(d) shows a larger scale image where such phase jumps occur in multiple places locally. This suggests that there are phase instabilities in the stripe order.
We find that these jumps are induced by tunneling electrons. We can track the phase jumps in time traces of the tunneling current (Fig.~\ref{stripedynamics}(e)). Fig.~\ref{stripedynamics}(f) shows histograms of current traces recorded over time windows from $60\mathrm{s}$ to $150\mathrm{s}$ as a function of bias, showing three different states between which the stripe order fluctuates. For bias voltages smaller than $20\mathrm{mV}$, no jumps were observed, but we consistently observe jumps for bias voltages from $20\mathrm{mV}$ up to $120\mathrm{mV}$. From the number of jumps seen in time traces as shown in Fig.~\ref{stripedynamics}(e), we can extract a switching frequency. The bias dependence of the switching frequency is shown in Fig.~\ref{stripedynamics}(g). It shows that the phase slips are only triggered by tunneling electrons above an energy threshold.
Analysis of the data shows that the minimum energy of the tunneling electrons required to trigger a phase slip is $20\mathrm{meV}$, and the process reaches a quantum yield $Y\sim 5\cdot 10^{-7}$ at $V=100\mathrm{mV}$.

\section{Discussion}
Stripe orders have been ubiquitous in cuprate superconductors, with evidence for fluctuating stripes in some of them, however stripe fluctuations or dynamics were never imaged in real space. Here, we show real space images of a stripe order and stripe dynamics in \ce{La4Ni3O10} in the non-superconducting ambient-pressure phase of a high-temperature nickelate superconductor. Our images of the charge and magnetic order reveal several aspects which are expected for a correlation-driven stripe order and which merit further experimental and theoretical studies: (1) From measurements of the Fermi surface \cite{li_fermiology_2017} and Raman spectroscopy \cite{suthar_multiorbital_2025} it has been argued that the charge order in \ce{La4Ni3O10} is driven by nesting \cite{li_fermiology_2017}, whereas our data suggests that the smallest characteristic wave vector of the magnetic order is $\mathbf{q}\sim 1/8$ which can not be linked to nesting. While nesting may very well play a role, the order is locally pinned and commensurate with the crystal lattice. This is supported both by the appearance of the stripe order characterized by multiple harmonics of the dominant $\mathbf{q}$-vector, and by commensurate stripe dynamics. This is consistent with a local moment order, as it occurs, e.g., due to stripes in the Mott picture. The wave vectors of the spin and charge order we observe are consistent with neutron scattering\cite{zhang_intertwined_2020}, apart from a small incommensurability seen in neutron scattering but not in our data which could be the result of regions with different commensurate arrangements (see suppl. Fig.~S4). The magnetic order accompanying the stripe modulations exhibits antiphase domains across the stripes, as would be predicted from a Mott Hubbard Hamiltonian. This is directly visible in our spin-polarized real space images. (2) The gap in tunneling spectroscopy is almost fully formed with only very small residual density of states. The tunneling spectrum can be rationalized by the $d_{z^2}$ band becoming fully gapped out, with only the $d_{x^2-y^2}$ band remaining at the Fermi energy but with low tunneling probability to the tip. Pure nesting scenarios would not be expected to gap out sufficiently large parts of the Fermi surface to account for such a near-complete depletion in the density of states. (3) Interesting open questions remain about the relation of the stripe physics predicted from a single-band Mott Hubbard Hamiltonian as is typically used for the cuprates translates to that of a multi-band system\cite{luo_charge_2011} as here where only one of the two bands is fully gapped out. The dynamics of the stripe order suggests that it is surprisingly weakly pinned and in fact support that with raising temperature, the order might undergo a transition to a fully fluctuating stripe order. Fluctuating stripes have been observed in high-temperature cuprate superconductors\cite{tranquada_evidence_1995,tranquada_coexistence_1997}, after their discovery in \ce{La2NiO4}\cite{hayden_incommensurate_1992,tranquada_simultaneous_1994}. While the fluctuations detected in neutron scattering occur on a much faster time scale, the fluctuations seen here may be a precursor to fluctuating order at higher doping levels as suggested for the cuprates. The relation of these fluctuations to the superconducting state has been the subject of intense research\cite{kivelson_electronic_1998}. Our data suggests strong similarities with the cuprates, putting the nickelates in the strongly correlated regime, consistent with their proximity to metal-to-insulator transitions (see, e.g., \cite{kuiper_unoccupied_1991,sanchez_metal-insulator_1996,kobayashi_transport_1996,sun_electronic_2021}). We expect that the real space imaging of a fluctuating stripe order will provide a new perspective onto the role of stripe orders in quantum critical phenomena in general\cite{kivelson_how_2003,sachdev_colloquium:_2003}, and specifically their relation with the Cooper pair condensate in high temperature superconductors.


\noindent{\bf Acknowledgements:} UM gratefully acknowledges support through the Grant family scholarship programme. LCR and PW acknowledge funding from the Leverhulme Trust through Research Project Grant RPG-2022-315, and PW from the Engineering and Physical Sciences Research Council through EP/X015556/1 and CDAM and PW from UKRI1107.  This work used computational resources of the Cirrus UK National Tier-2 HPC Service at EPCC (http://www.cirrus.ac.uk) funded by the University of Edinburgh and EPSRC (EP/P020267/1) and Archer2 in Edinburgh.\\

\noindent{\bf Author Contributions:} 
UM did STM measurements with assistance of LCR, UM, CDAM and PW did spin-polarized measurements, LCR and UM did calculations, SAB prepared and characterized the field-emission target for spin-polarized measurements, MI and PP grew and characterized the samples. UM, LCR and PW wrote the manuscript. PW initiated and supervised the project. All authors discussed and contributed to the manuscript.\\
\noindent{\bf Competing Interests:} The authors declare that they have no competing interests.\\

\section{Methods}
\subsection{Single crystal growth}
For the optical floating zone (OFZ) growth of La$_4$Ni$_3$O$_{10-x}$,  we dried La$_2$O$_3$ powder (99.99\% Alfa Aesar) in a box furnace at 1100$^\circ$C. Subsequently, the precursor powder was prepared by mixing La$_2$O$_3$ and NiO (99.998\% Alfa Aesar) according to a 4:3.05 stoichiometry of La:Ni. The additional NiO is added to cover for the NiO loss from evaporation and stay away from the eutectic. The mixture was ball-milled for 20 min and transferred in alumina crucibles to a box furnace, followed by heating to 1100$^{\circ}$C for 24 hours.
Cylindrically shaped feed and seed rods were prepared by ball-milling of the sintered materials, which were filled into rubber forms with 6~mm diameter. The rubber was evacuated and pressed in a stainless steel form filled with water using a Riken type S1-120 70 kN press. All rods were heat treated at 1150$^{\circ}$C.

The single-crystal growth was carried out in a high pressure, high-temperature OFZ furnace (model HKZ, SciDre GmbH, Dresden, Germany), that allows for gas pressures in the growth chamber up to 300 bar. The growth chamber (sapphire single crystal) has a length of 72~mm and a wall thickness of 20~mm. A Xe arc lamp operating at 5 kW was used as a heating source within the vertical mirror alignment of the HKZ. The 14 cm feed and 4 cm seed rods were then aligned in the HKZ on steel holders followed by the installation of the high pressure chamber. Subsequently, for the growth the chamber was pressurized with 20 bar oxygen gas and held at a flow rate of 0.1 l/min. After connecting the molten zone, the growth was carried out by moving the seed with a speed of 4 mm/h and the feed at 11 mm/h. The boule was inspected with XRD and several mm sized single crystals were extracted. 

High pressure annealing was done by flushing an autoclave several times with oxygen and subsequently filling it with 200 bar oxygen pressure. The pressure was increased with a Nova swiss membrane compressor to 500 bar. The small sample volume was then heated to 600\degree C leading to an increase of the pressure to 600 bar. The pressure and temperature was held for 5 days and subsequently quenched to room temperature, where the pressure was slowly released.

\subsection{Scanning Tunneling Microscopy}
STM measurements were performed using a home-built low temperature STM mounted in a vector magnet\cite{trainer_cryogenic_2017} with a head made out of sapphire\cite{white_stiff_2011}. The STM routinely operates at a temperature of $1.8\mathrm{K}$. The bias voltage is applied to the sample with the tip at virtual ground. Tunneling spectra were recorded using the usual lock-in technique. Samples were cleaved in-situ at a cleaving stage at the $4\mathrm{K}$-plate.

\subsection{Spin-polarized STM measurements}
For spin-polarized measurements, we have prepared the tip on an iron field emission target coated by a thin gold film to prevent oxidation. The field emission target was created by cutting a high purity iron foil with a thickness of 0.5 mm and a purity of 99.5\% into a $2\times 2 \mathrm{mm}^2$ target. The surface of the target was cleaned by Argon ion etching followed by deposition of 18nm of gold using an Angstrom Engineering Ion-assisted e-beam evaporator. Characterization by transmission electron microscopy confirmed the film thickness. EDX shows no signs of oxidation at the interface between the Fe foil and the gold layer.

\label{Bibliography}
\bibliographystyle{unsrtnat}
\bibliography{la4ni3o10}

\begin{thebibliography}{42}
\providecommand{\natexlab}[1]{#1}
\providecommand{\url}[1]{\texttt{#1}}
\expandafter\ifx\csname urlstyle\endcsname\relax
  \providecommand{\doi}[1]{doi: #1}\else
  \providecommand{\doi}{doi: \begingroup \urlstyle{rm}\Url}\fi

\bibitem[Sun et~al.(2023)Sun, Huo, Hu, Li, Liu, Han, Tang, Mao, Yang, Wang,
  Cheng, Yao, Zhang, and Wang]{sun_signatures_2023}
Hualei Sun, Mengwu Huo, Xunwu Hu, Jingyuan Li, Zengjia Liu, Yifeng Han, Lingyun
  Tang, Zhongquan Mao, Pengtao Yang, Bosen Wang, Jinguang Cheng, Dao-Xin Yao,
  Guang-Ming Zhang, and Meng Wang.
\newblock Signatures of superconductivity near 80 {K} in a nickelate under high
  pressure.
\newblock \emph{Nature}, 621:\penalty0 493--498, 2023.
\newblock \doi{10.1038/s41586-023-06408-7}.
\newblock URL \url{https://www.nature.com/articles/s41586-023-06408-7}.

\bibitem[Zaanen and Gunnarsson(1989)]{zaanen_charged_1989}
Jan Zaanen and Olle Gunnarsson.
\newblock Charged magnetic domain lines and the magnetism of
  high-{$T_\mathrm{c}$} oxides.
\newblock \emph{Phys. Rev. B}, 40:\penalty0 7391--7394, October 1989.
\newblock \doi{10.1103/PhysRevB.40.7391}.
\newblock URL \url{https://link.aps.org/doi/10.1103/PhysRevB.40.7391}.

\bibitem[Kivelson et~al.(1998)Kivelson, Fradkin, and
  Emery]{kivelson_electronic_1998}
S.~A. Kivelson, E.~Fradkin, and V.~J. Emery.
\newblock Electronic liquid-crystal phases of a doped {Mott} insulator.
\newblock \emph{Nature}, 393:\penalty0 550--553, 1998.
\newblock \doi{10.1038/31177}.
\newblock URL \url{http://www.nature.com/articles/31177}.

\bibitem[Tranquada et~al.(1995{\natexlab{a}})Tranquada, Lorenzo, Buttrey, and
  Sachan]{tranquada_cooperative_1995}
J.~M. Tranquada, J.~E. Lorenzo, D.~J. Buttrey, and V.~Sachan.
\newblock Cooperative ordering of holes and spins in
  {La}$_{\textrm{2}}${NiO}$_{\textrm{4.125}}$.
\newblock \emph{Phys. Rev. B}, 52:\penalty0 3581--3595, 1995{\natexlab{a}}.
\newblock ISSN 0163-1829, 1095-3795.
\newblock \doi{10.1103/PhysRevB.52.3581}.
\newblock URL \url{https://link.aps.org/doi/10.1103/PhysRevB.52.3581}.

\bibitem[Tranquada et~al.(1997)Tranquada, Axe, Ichikawa, Moodenbaugh, Nakamura,
  and Uchida]{tranquada_coexistence_1997}
J.~M. Tranquada, J.~D. Axe, N.~Ichikawa, A.~R. Moodenbaugh, Y.~Nakamura, and
  S.~Uchida.
\newblock Coexistence of, and {Competition} between, {Superconductivity} and
  {Charge}-{Stripe} {Order} in \ce{La_{1.6-x}Nd_{0.4}Sr_xCuO4}.
\newblock \emph{Physical Review Letters}, 78:\penalty0 338--341, January 1997.
\newblock \doi{10.1103/PhysRevLett.78.338}.
\newblock URL \url{https://link.aps.org/doi/10.1103/PhysRevLett.78.338}.

\bibitem[Tranquada et~al.(2008)Tranquada, Gu, Hücker, Jie, Kang, Klingeler,
  Li, Tristan, Wen, Xu, Xu, Zhou, and V.~Zimmermann]{tranquada_evidence_2008}
J.~M. Tranquada, G.~D. Gu, M.~Hücker, Q.~Jie, H.-J. Kang, R.~Klingeler, Q.~Li,
  N.~Tristan, J.~S. Wen, G.~Y. Xu, Z.~J. Xu, J.~Zhou, and M.~V.~Zimmermann.
\newblock Evidence for unusual superconducting correlations coexisting with
  stripe order in \ce{La_{1.875}Ba_{0.125}CuO_4}.
\newblock \emph{Phys. Rev. B}, 78:\penalty0 174529, 2008.
\newblock \doi{10.1103/PhysRevB.78.174529}.
\newblock URL \url{https://link.aps.org/doi/10.1103/PhysRevB.78.174529}.

\bibitem[Ghiringhelli et~al.(2012)Ghiringhelli, Le~Tacon, Minola,
  Blanco-Canosa, Mazzoli, Brookes, De~Luca, Frano, Hawthorn, He, Loew, Sala,
  Peets, Salluzzo, Schierle, Sutarto, Sawatzky, Weschke, Keimer, and
  Braicovich]{ghiringhelli_long-range_2012}
G.~Ghiringhelli, M.~Le~Tacon, M.~Minola, S.~Blanco-Canosa, C.~Mazzoli, N.~B.
  Brookes, G.~M. De~Luca, A.~Frano, D.~G. Hawthorn, F.~He, T.~Loew, M.~M. Sala,
  D.~C. Peets, M.~Salluzzo, E.~Schierle, R.~Sutarto, G.~A. Sawatzky,
  E.~Weschke, B.~Keimer, and L.~Braicovich.
\newblock Long-{Range} {Incommensurate} {Charge} {Fluctuations} in
  \ce{({Y},{Nd})Ba2Cu3O_{6+x}}.
\newblock \emph{Science}, 337:\penalty0 821--825, 2012.
\newblock \doi{10.1126/science.1223532}.
\newblock URL \url{http://www.sciencemag.org/cgi/doi/10.1126/science.1223532}.

\bibitem[Emery and Kivelson(1995)]{emery_importance_1995}
V.~J. Emery and S.~A. Kivelson.
\newblock Importance of phase fluctuations in superconductors with small
  superfluid density.
\newblock \emph{Nature}, 374:\penalty0 434--437, 1995.
\newblock \doi{10.1038/374434a0}.
\newblock URL \url{https://www.nature.com/articles/374434a0}.

\bibitem[Loder et~al.(2011)Loder, Graser, Schmid, Kampf, and
  Kopp]{loder_modeling_2011}
F~Loder, S~Graser, M~Schmid, A~P Kampf, and T~Kopp.
\newblock Modeling of superconducting stripe phases in high-$t_c$ cuprates.
\newblock \emph{New J. Phys.}, 13:\penalty0 113037, 2011.
\newblock \doi{10.1088/1367-2630/13/11/113037}.
\newblock URL
  \url{https://iopscience.iop.org/article/10.1088/1367-2630/13/11/113037}.

\bibitem[Jiang and Kivelson(2022)]{jiang_stripe_2022}
Hong-Chen Jiang and Steven~A. Kivelson.
\newblock Stripe order enhanced superconductivity in the {Hubbard} model.
\newblock \emph{Proc. Natl. Acad. Sci. U.S.A.}, 119:\penalty0 e2109406119,
  2022.
\newblock \doi{10.1073/pnas.2109406119}.
\newblock URL \url{https://pnas.org/doi/full/10.1073/pnas.2109406119}.

\bibitem[Hücker et~al.(2007)Hücker, Gu, Tranquada, Zimmermann, Klauss, Curro,
  Braden, and Büchner]{hucker_coupling_2007}
M.~Hücker, G.D. Gu, J.M. Tranquada, M.v. Zimmermann, H.-H. Klauss, N.J. Curro,
  M.~Braden, and B.~Büchner.
\newblock Coupling of stripes to lattice distortions in cuprates and
  nickelates.
\newblock \emph{Physica C: Superconductivity and its Applications},
  460-462:\penalty0 170--173, 2007.
\newblock \doi{https://doi.org/10.1016/j.physc.2007.03.003}.
\newblock URL
  \url{https://www.sciencedirect.com/science/article/pii/S092145340700072X}.

\bibitem[Hanaguri et~al.(2004)Hanaguri, Lupien, Kohsaka, Lee, Azuma, Takano,
  Takagi, and Davis]{hanaguri_checkerboard_2004}
T.~Hanaguri, C.~Lupien, Y.~Kohsaka, D.-H. Lee, M.~Azuma, M.~Takano, H.~Takagi,
  and J.~C. Davis.
\newblock A ‘checkerboard’ electronic crystal state in lightly hole-doped
  \ce{Ca_{2-x}Na_xCuO_2Cl_2}.
\newblock \emph{Nature}, 430:\penalty0 1001--1005, 2004.
\newblock \doi{10.1038/nature02861}.
\newblock URL \url{http://www.nature.com/doifinder/10.1038/nature02861}.

\bibitem[Kohsaka et~al.(2007)Kohsaka, Taylor, Fujita, Schmidt, Lupien,
  Hanaguri, Azuma, Takano, Eisaki, Takagi, Uchida, and
  Davis]{kohsaka_intrinsic_2007}
Y.~Kohsaka, C.~Taylor, K.~Fujita, A.~Schmidt, C.~Lupien, T.~Hanaguri, M.~Azuma,
  M.~Takano, H.~Eisaki, H.~Takagi, S.~Uchida, and J.~C. Davis.
\newblock An {Intrinsic} {Bond}-{Centered} {Electronic} {Glass} with
  {Unidirectional} {Domains} in {Underdoped} {Cuprates}.
\newblock \emph{Science}, 315:\penalty0 1380--1385, 2007.
\newblock \doi{10.1126/science.1138584}.
\newblock URL \url{http://www.sciencemag.org/cgi/doi/10.1126/science.1138584}.

\bibitem[Comin et~al.(2014)Comin, Frano, Yee, Yoshida, Eisaki, Schierle,
  Weschke, Sutarto, He, Soumyanarayanan, He, Le~Tacon, Elfimov, Hoffman,
  Sawatzky, Keimer, and Damascelli]{comin_charge_2014}
R.~Comin, A.~Frano, M.~M. Yee, Y.~Yoshida, H.~Eisaki, E.~Schierle, E.~Weschke,
  R.~Sutarto, F.~He, A.~Soumyanarayanan, Y.~He, M.~Le~Tacon, I.~S. Elfimov,
  J.~E. Hoffman, G.~A. Sawatzky, B.~Keimer, and A.~Damascelli.
\newblock Charge {Order} {Driven} by {Fermi}-{Arc} {Instability} in
  \ce{Bi2Sr_{2-x}La_xCuO_{6+\delta}}.
\newblock \emph{Science}, 343:\penalty0 390--392, 2014.
\newblock \doi{10.1126/science.1242996}.
\newblock URL \url{http://www.sciencemag.org/cgi/doi/10.1126/science.1242996}.

\bibitem[Wang et~al.(2024)Wang, Wang, Shen, Hou, Ma, Shi, Ren, Gu, Ma, Yang,
  Liu, Guo, Sun, Zhang, Calder, Yan, Wang, Uwatoko, and
  Cheng]{wang_pressure-induced_2024}
G.~Wang, N.N. Wang, X.L. Shen, J.~Hou, L.~Ma, L.F. Shi, Z.A. Ren, Y.D. Gu, H.M.
  Ma, P.T. Yang, Z.Y. Liu, H.Z. Guo, J.P. Sun, G.M. Zhang, S.~Calder, J.-Q.
  Yan, B.S. Wang, Y.~Uwatoko, and J.-G. Cheng.
\newblock Pressure-{Induced} {Superconductivity} {In} {Polycrystalline}
  \ce{La3Ni2O_{7-\delta}}.
\newblock \emph{Phys. Rev. X}, 14:\penalty0 011040, 2024.
\newblock \doi{10.1103/PhysRevX.14.011040}.
\newblock URL \url{https://link.aps.org/doi/10.1103/PhysRevX.14.011040}.

\bibitem[Zhu et~al.(2024)Zhu, Peng, Zhang, Pan, Chen, Chen, Ren, Liu, Hao, Li,
  Xing, Lan, Han, Wang, Jia, Wo, Gu, Gu, Ji, Wang, Gou, Shen, Ying, Chen, Yang,
  Cao, Zheng, Zeng, Guo, and Zhao]{zhu_superconductivity_2024}
Yinghao Zhu, Di~Peng, Enkang Zhang, Bingying Pan, Xu~Chen, Lixing Chen, Huifen
  Ren, Feiyang Liu, Yiqing Hao, Nana Li, Zhenfang Xing, Fujun Lan, Jiyuan Han,
  Junjie Wang, Donghan Jia, Hongliang Wo, Yiqing Gu, Yimeng Gu, Li~Ji, Wenbin
  Wang, Huiyang Gou, Yao Shen, Tianping Ying, Xiaolong Chen, Wenge Yang, Huibo
  Cao, Changlin Zheng, Qiaoshi Zeng, Jian-gang Guo, and Jun Zhao.
\newblock Superconductivity in pressurized trilayer \ce{La4Ni3O_{10-\delta}}
  single crystals.
\newblock \emph{Nature}, 631:\penalty0 531--536, 2024.
\newblock \doi{10.1038/s41586-024-07553-3}.
\newblock URL \url{https://www.nature.com/articles/s41586-024-07553-3}.

\bibitem[Zhang et~al.(2020)Zhang, Phelan, Botana, Chen, Zheng, Krogstad, Wang,
  Qiu, Rodriguez-Rivera, Osborn, Rosenkranz, Norman, and
  Mitchell]{zhang_intertwined_2020}
Junjie Zhang, D.~Phelan, A.~S. Botana, Yu-Sheng Chen, Hong Zheng, M.~Krogstad,
  Suyin~Grass Wang, Yiming Qiu, J.~A. Rodriguez-Rivera, R.~Osborn,
  S.~Rosenkranz, M.~R. Norman, and J.~F. Mitchell.
\newblock Intertwined density waves in a metallic nickelate.
\newblock \emph{Nat Commun}, 11:\penalty0 6003, 2020.
\newblock \doi{10.1038/s41467-020-19836-0}.
\newblock URL \url{https://www.nature.com/articles/s41467-020-19836-0}.

\bibitem[Kobayashi et~al.(1996)Kobayashi, Taniguchi, Kasai, Sato, Nishioka, and
  Kontani]{kobayashi_transport_1996}
Yoshiaki Kobayashi, Satoshi Taniguchi, Mayumi Kasai, Masatoshi Sato, Takashi
  Nishioka, and Masaaki Kontani.
\newblock Transport and {Magnetic} {Properties} of \ce{La_3Ni_2O_{7-\delta}}
  and \ce{La_4Ni_3O_{10-\delta}}.
\newblock \emph{J. Phys. Soc. Jpn.}, 65:\penalty0 3978--3982, 1996.
\newblock \doi{10.1143/JPSJ.65.3978}.
\newblock URL \url{http://journals.jps.jp/doi/10.1143/JPSJ.65.3978}.

\bibitem[Yang et~al.(2024)Yang, Sun, Hu, Xie, Miao, Luo, Chen, Liang, Zhu, Qu,
  Chen, Huo, Huang, Zhang, Zhang, Yang, Wang, Peng, Mao, Liu, Xu, Qian, Yao,
  Wang, Zhao, and Zhou]{yang_orbital-dependent_2024}
Jiangang Yang, Hualei Sun, Xunwu Hu, Yuyang Xie, Taimin Miao, Hailan Luo, Hao
  Chen, Bo~Liang, Wenpei Zhu, Gexing Qu, Cui-Qun Chen, Mengwu Huo, Yaobo Huang,
  Shenjin Zhang, Fengfeng Zhang, Feng Yang, Zhimin Wang, Qinjun Peng, Hanqing
  Mao, Guodong Liu, Zuyan Xu, Tian Qian, Dao-Xin Yao, Meng Wang, Lin Zhao, and
  X.~J. Zhou.
\newblock Orbital-dependent electron correlation in double-layer nickelate
  \ce{La3Ni2O7}.
\newblock \emph{Nat Commun}, 15:\penalty0 4373, 2024.
\newblock \doi{10.1038/s41467-024-48701-7}.
\newblock URL \url{https://www.nature.com/articles/s41467-024-48701-7}.

\bibitem[Li et~al.(2026)Li, Zhang, Du, Pei, Liu, Chen, Zhao, Hu, Zhang, Shao,
  Mao, Cao, Zhao, Li, Shen, Huang, Hashimoto, Lu, Liu, Chen, Wang, Qi, and
  Yang]{li_orbital-selective_nodate}
Yidian Li, Mingxin Zhang, Xian Du, Cuiying Pei, Jieyi Liu, Houke Chen, Wenxuan
  Zhao, Yinqi Hu, Senyao Zhang, Jiawei Shao, Mingxin Mao, Yantao Cao, Jinkui
  Zhao, Zhengtai Li, Dawei Shen, Yaobo Huang, Makoto Hashimoto, Donghui Lu,
  Zhongkai Liu, Yulin Chen, Yilin Wang, Yanpeng Qi, and Lexian Yang.
\newblock Orbital-selective {Mottness} {Driven} by {Geometric} {Frustration} of
  {Interorbital} {Hybridization} in \ce{Pr4Ni3O10}, 2026.
\newblock URL \url{http://arxiv.org/abs/2602.03658}.
\newblock arXiv:2602.03658 [cond-mat].

\bibitem[Sohn et~al.(2025)Sohn, Kim, Lee, Wei, Jiang, Li, Gorovikov, Zonno,
  Pedersen, Zhdanovich, Liu, Cheng, Zou, He, Ismail-Beigi, Walker, and
  Ahn]{sohn_layer-controlled_2025}
Byungmin Sohn, Minjae Kim, Sangjae Lee, Wenzheng Wei, Juan Jiang, Fengmiao Li,
  Sergey Gorovikov, Marta Zonno, Tor Pedersen, Sergey Zhdanovich, Ying Liu,
  Huikai Cheng, Ke~Zou, Yu~He, Sohrab Ismail-Beigi, Frederick~J. Walker, and
  Charles~H. Ahn.
\newblock Layer-controlled orbital-selective {Mott} transition in monolayer
  nickelate.
\newblock \emph{Phys. Rev. Research}, 7:\penalty0 043132, 2025.
\newblock \doi{10.1103/n92p-xrhl}.
\newblock URL \url{https://link.aps.org/doi/10.1103/n92p-xrhl}.

\bibitem[Hudson et~al.(2001)Hudson, Lang, Madhavan, Pan, Eisaki, Uchida, and
  Davis]{hudson_interplay_2001}
E.~W. Hudson, K.~M. Lang, V.~Madhavan, S.~H. Pan, H.~Eisaki, S.~Uchida, and
  J.~C. Davis.
\newblock Interplay of magnetism and high-{Tc} superconductivity at individual
  {Ni} impurity atoms in \ce{Bi2Sr2CaCu2O_{8+\delta}}.
\newblock \emph{Nature}, 411:\penalty0 920--924, 2001.
\newblock \doi{10.1038/35082019}.
\newblock URL
  \url{http://www.nature.com/nature/journal/v411/n6840/full/411920a0.html}.

\bibitem[Marques et~al.(2021)Marques, Rhodes, Fittipaldi, Granata, Yim, Buzio,
  Gerbi, Vecchione, Rost, and Wahl]{marques_magneticfield_2021}
Carolina~A Marques, Luke~C Rhodes, Rosalba Fittipaldi, Veronica Granata,
  Chi~Ming Yim, Renato Buzio, Andrea Gerbi, Antonio Vecchione, Andreas~W Rost,
  and Peter Wahl.
\newblock Magnetic‐{Field} {Tunable} {Intertwined} {Checkerboard} {Charge}
  {Order} and {Nematicity} in the {Surface} {Layer} of \ce{Sr2RuO4}.
\newblock \emph{Advanced Materials}, 33:\penalty0 2100593, 2021.
\newblock \doi{10.1002/adma.202100593}.
\newblock URL \url{https://onlinelibrary.wiley.com/doi/10.1002/adma.202100593}.

\bibitem[Li et~al.(2025)Li, Gong, Zhu, Chen, Zhang, Zhang, Li, Yin, Wang, Zhao,
  Feng, Du, and Yan]{li_direct_2025}
Mingzhe Li, Jiashuo Gong, Yinghao Zhu, Ziyuan Chen, Jiakang Zhang, Enkang
  Zhang, Yuanji Li, Ruotong Yin, Shiyuan Wang, Jun Zhao, Dong-Lai Feng, Zengyi
  Du, and Ya-Jun Yan.
\newblock Direct visualization of an incommensurate unidirectional charge
  density wave in \ce{La4Ni3O10}.
\newblock \emph{Phys. Rev. B}, 112:\penalty0 045132, 2025.
\newblock \doi{10.1103/2p56-xl41}.
\newblock URL \url{https://link.aps.org/doi/10.1103/2p56-xl41}.

\bibitem[Halwidl et~al.(2017)Halwidl, Mayr-Schmölzer, Fobes, Peng, Mao,
  Schmid, Mittendorfer, Redinger, and Diebold]{halwidl_ordered_2017}
Daniel Halwidl, Wernfried Mayr-Schmölzer, David Fobes, Jin Peng, Zhiqiang Mao,
  Michael Schmid, Florian Mittendorfer, Josef Redinger, and Ulrike Diebold.
\newblock Ordered hydroxyls on \ce{Ca3Ru2O7}(001).
\newblock \emph{Nat Commun}, 8:\penalty0 23, 2017.
\newblock \doi{10.1038/s41467-017-00066-w}.
\newblock URL \url{https://www.nature.com/articles/s41467-017-00066-w}.

\bibitem[Anderson and Ong(2006)]{anderson_theory_2006}
P.W. Anderson and N.P. Ong.
\newblock Theory of asymmetric tunneling in the cuprate superconductors.
\newblock \emph{Journal of Physics and Chemistry of Solids}, 67\penalty0
  (1-3):\penalty0 1--5, 2006.
\newblock \doi{10.1016/j.jpcs.2005.10.132}.
\newblock URL
  \url{https://linkinghub.elsevier.com/retrieve/pii/S0022369705004063}.

\bibitem[Periyasamy et~al.(2021)Periyasamy, Patra, Fjellvåg, Ravindran,
  Sørby, Kumar, Sjåstad, and Fjellvåg]{periyasamy_effect_2021}
Manimuthu Periyasamy, Lokanath Patra, Øystein~S. Fjellvåg, Ponniah Ravindran,
  Magnus~H. Sørby, Susmit Kumar, Anja~O. Sjåstad, and Helmer Fjellvåg.
\newblock Effect of {Electron} {Doping} on the {Crystal} {Structure} and
  {Physical} {Properties} of an $n = 3$ {Ruddlesden}–{Popper} {Compound}
  \ce{La4Ni3O10}.
\newblock \emph{ACS Appl. Electron. Mater.}, 3:\penalty0 2671--2684, 2021.
\newblock \doi{10.1021/acsaelm.1c00270}.
\newblock URL \url{https://pubs.acs.org/doi/10.1021/acsaelm.1c00270}.

\bibitem[Wiesendanger(2009)]{wiesendanger_spin_2009}
Roland Wiesendanger.
\newblock Spin mapping at the nanoscale and atomic scale.
\newblock \emph{Reviews of Modern Physics}, 81:\penalty0 1495--1550, 2009.
\newblock \doi{10.1103/RevModPhys.81.1495}.
\newblock URL \url{http://link.aps.org/doi/10.1103/RevModPhys.81.1495}.

\bibitem[Enayat et~al.(2014)Enayat, Sun, Singh, Aluru, Schmaus, Yaresko, Liu,
  Lin, Tsurkan, Loidl, Deisenhofer, and Wahl]{enayat_real-space_2014}
M.~Enayat, Z.~Sun, U.~R. Singh, R.~Aluru, S.~Schmaus, A.~Yaresko, Y.~Liu,
  C.~Lin, V.~Tsurkan, A.~Loidl, J.~Deisenhofer, and P.~Wahl.
\newblock Real-space imaging of the atomic-scale magnetic structure of
  {Fe}$_{\textrm{1+\textit{y}}}${Te}.
\newblock \emph{Science}, 345:\penalty0 653--656, 2014.
\newblock \doi{10.1126/science.1251682}.
\newblock URL \url{http://www.sciencemag.org/cgi/doi/10.1126/science.1251682}.

\bibitem[Li et~al.(2017)Li, Zhou, Nummy, Zhang, Pardo, Pickett, Mitchell, and
  Dessau]{li_fermiology_2017}
Haoxiang Li, Xiaoqing Zhou, Thomas Nummy, Junjie Zhang, Victor Pardo, Warren~E.
  Pickett, J.~F. Mitchell, and D.~S. Dessau.
\newblock Fermiology and electron dynamics of trilayer nickelate
  \ce{La4Ni3O10}.
\newblock \emph{Nat Commun}, 8:\penalty0 704, 2017.
\newblock \doi{10.1038/s41467-017-00777-0}.
\newblock URL \url{https://www.nature.com/articles/s41467-017-00777-0}.

\bibitem[Suthar et~al.(2025)Suthar, Sundaramurthy, Bejas, Le, Puphal,
  Sosa-Lizama, Schulz, Nuss, Isobe, Aken, Suyolcu, Minola, Schnyder, Wu,
  Keimer, Khaliullin, Greco, and Hepting]{suthar_multiorbital_2025}
A.~Suthar, V.~Sundaramurthy, M.~Bejas, Congcong Le, P.~Puphal, P.~Sosa-Lizama,
  A.~Schulz, J.~Nuss, M.~Isobe, P.~A.~van Aken, Y.~E. Suyolcu, M.~Minola, A.~P.
  Schnyder, Xianxin Wu, B.~Keimer, G.~Khaliullin, A.~Greco, and M.~Hepting.
\newblock Multiorbital character of the density wave instability in
  \ce{La4Ni3O_{10}}, 2025.
\newblock URL \url{http://arxiv.org/abs/2508.06440}.
\newblock arXiv:2508.06440 [cond-mat].

\bibitem[Luo et~al.(2011)Luo, Yao, Moreo, and Dagotto]{luo_charge_2011}
Qinlong Luo, Dao-Xin Yao, Adriana Moreo, and Elbio Dagotto.
\newblock Charge stripes in the two-orbital {Hubbard} model for iron pnictides.
\newblock \emph{Phys. Rev. B}, 83:\penalty0 174513, 2011.
\newblock \doi{10.1103/PhysRevB.83.174513}.
\newblock URL \url{https://link.aps.org/doi/10.1103/PhysRevB.83.174513}.

\bibitem[Tranquada et~al.(1995{\natexlab{b}})Tranquada, Sternlieb, Axe,
  Nakamura, and Uchida]{tranquada_evidence_1995}
J.~M. Tranquada, B.~J. Sternlieb, J.~D. Axe, Y.~Nakamura, and S.~Uchida.
\newblock Evidence for stripe correlations of spins and holes in copper oxide
  superconductors.
\newblock \emph{Nature}, 375:\penalty0 561--563, 1995{\natexlab{b}}.
\newblock \doi{10.1038/375561a0}.
\newblock URL \url{http://www.nature.com/doifinder/10.1038/375561a0}.

\bibitem[Hayden et~al.(1992)Hayden, Lander, Zarestky, Brown, Stassis, Metcalf,
  and Honig]{hayden_incommensurate_1992}
S.~M. Hayden, G.~H. Lander, J.~Zarestky, P.~J. Brown, C.~Stassis, P.~Metcalf,
  and J.~M. Honig.
\newblock Incommensurate magnetic correlations in \ce{La_{1.8}Sr_{0.2}NiO_4}.
\newblock \emph{Phys. Rev. Lett.}, 68:\penalty0 1061--1064, 1992.
\newblock \doi{10.1103/PhysRevLett.68.1061}.
\newblock URL \url{https://link.aps.org/doi/10.1103/PhysRevLett.68.1061}.

\bibitem[Tranquada et~al.(1994)Tranquada, Buttrey, Sachan, and
  Lorenzo]{tranquada_simultaneous_1994}
J.~M. Tranquada, D.~J. Buttrey, V.~Sachan, and J.~E. Lorenzo.
\newblock Simultaneous {Ordering} of {Holes} and {Spins} in
  \ce{La_2NiO_{4.125}}.
\newblock \emph{Phys. Rev. Lett.}, 73:\penalty0 1003--1006, 1994.
\newblock \doi{10.1103/PhysRevLett.73.1003}.
\newblock URL \url{https://link.aps.org/doi/10.1103/PhysRevLett.73.1003}.

\bibitem[Kuiper et~al.(1991)Kuiper, Van~Elp, Sawatzky, Fujimori, Hosoya, and
  De~Leeuw]{kuiper_unoccupied_1991}
P.~Kuiper, J.~Van~Elp, G.~A. Sawatzky, A.~Fujimori, S.~Hosoya, and D.~M.
  De~Leeuw.
\newblock Unoccupied density of states of \ce{La_{2-x}{Sr}_xNiO_{4+\delta}}
  studied by polarization-dependent x-ray-absorption spectroscopy and
  bremsstrahlung isochromat spectroscopy.
\newblock \emph{Phys. Rev. B}, 44:\penalty0 4570--4575, 1991.
\newblock \doi{10.1103/PhysRevB.44.4570}.
\newblock URL \url{https://link.aps.org/doi/10.1103/PhysRevB.44.4570}.

\bibitem[Sánchez et~al.(1996)Sánchez, Causa, Caneiro, Butera, Vallet-Regí,
  Sayagués, González-Calbet, García-Sanz, and
  Rivas]{sanchez_metal-insulator_1996}
R.~D. Sánchez, M.~T. Causa, A.~Caneiro, A.~Butera, M.~Vallet-Regí, M.~J.
  Sayagués, J.~González-Calbet, F.~García-Sanz, and J.~Rivas.
\newblock Metal-insulator transition in oxygen-deficient \ce{LaNiO_{3-x}}
  perovskites.
\newblock \emph{Phys. Rev. B}, 54:\penalty0 16574--16578, 1996.
\newblock \doi{10.1103/PhysRevB.54.16574}.
\newblock URL \url{https://link.aps.org/doi/10.1103/PhysRevB.54.16574}.

\bibitem[Sun et~al.(2021)Sun, Li, Cai, Yang, Guo, Gu, Zhu, and
  Nie]{sun_electronic_2021}
Wenjie Sun, Yueying Li, Xiangbin Cai, Jiangfeng Yang, Wei Guo, Zhengbin Gu,
  Ye~Zhu, and Yuefeng Nie.
\newblock Electronic and transport properties in {Ruddlesden}-{Popper}
  neodymium nickelates \ce{Nd_{n+1}Ni_nO_{3n+1}} ($n=1-5$).
\newblock \emph{Phys. Rev. B}, 104:\penalty0 184518, 2021.
\newblock \doi{10.1103/PhysRevB.104.184518}.
\newblock URL \url{https://link.aps.org/doi/10.1103/PhysRevB.104.184518}.

\bibitem[Kivelson et~al.(2003)Kivelson, Bindloss, Fradkin, Oganesyan,
  Tranquada, Kapitulnik, and Howald]{kivelson_how_2003}
Steven~A. Kivelson, I.~P. Bindloss, E.~Fradkin, V.~Oganesyan, J.~M. Tranquada,
  Aharon Kapitulnik, and Craig Howald.
\newblock How to detect fluctuating stripes in the high-temperature
  superconductors.
\newblock \emph{Reviews of Modern Physics}, 75:\penalty0 1201, 2003.
\newblock URL
  \url{http://journals.aps.org/rmp/abstract/10.1103/RevModPhys.75.1201}.

\bibitem[Sachdev(2003)]{sachdev_colloquium:_2003}
Subir Sachdev.
\newblock Colloquium: {Order} and quantum phase transitions in the cuprate
  superconductors.
\newblock \emph{Reviews of Modern Physics}, 75:\penalty0 913, 2003.
\newblock URL
  \url{http://journals.aps.org/rmp/abstract/10.1103/RevModPhys.75.913}.

\bibitem[Trainer et~al.(2017)Trainer, Yim, McLaren, and
  Wahl]{trainer_cryogenic_2017}
C.~Trainer, C.~M. Yim, M.~McLaren, and P.~Wahl.
\newblock Cryogenic {STM} in {3D} vector magnetic fields realized through a
  rotatable insert.
\newblock \emph{Review of Scientific Instruments}, 88:\penalty0 093705, 2017.
\newblock \doi{10.1063/1.4995688}.
\newblock URL \url{http://aip.scitation.org/doi/10.1063/1.4995688}.

\bibitem[White et~al.(2011)White, Singh, and Wahl]{white_stiff_2011}
S.~C. White, U.~R. Singh, and P.~Wahl.
\newblock A stiff scanning tunneling microscopy head for measurement at low
  temperatures and in high magnetic fields.
\newblock \emph{Review of Scientific Instruments}, 82:\penalty0 113708, 2011.
\newblock URL
  \url{http://scitation.aip.org/content/aip/journal/rsi/82/11/10.1063/1.3663611}.

\end{thebibliography}


\begin{thebibliography}{11}
\providecommand{\natexlab}[1]{#1}
\providecommand{\url}[1]{\texttt{#1}}
\expandafter\ifx\csname urlstyle\endcsname\relax
  \providecommand{\doi}[1]{doi: #1}\else
  \providecommand{\doi}{doi: \begingroup \urlstyle{rm}\Url}\fi

\bibitem[Anderson and Ong(2006)]{anderson_theory_2006}
P.W. Anderson and N.P. Ong.
\newblock Theory of asymmetric tunneling in the cuprate superconductors.
\newblock \emph{Journal of Physics and Chemistry of Solids}, 67\penalty0
  (1-3):\penalty0 1--5, 2006.
\newblock \doi{10.1016/j.jpcs.2005.10.132}.
\newblock URL
  \url{https://linkinghub.elsevier.com/retrieve/pii/S0022369705004063}.

\bibitem[Wahl et~al.(2025)Wahl, Rhodes, and Marques]{wahl_calcqpi_2025}
Peter Wahl, Luke~C Rhodes, and Carolina~A Marques.
\newblock {calcQPI}: {A} versatile tool to simulate quasiparticle interference.
\newblock \emph{SciPost Physics Codebases}, page~61, 2025.
\newblock \doi{10.21468/SciPostPhysCodeb.61}.

\bibitem[Kresse and Hafner(1993)]{kresse_ab_1993}
G.~Kresse and J.~Hafner.
\newblock \textit{{Ab} initio} molecular dynamics for liquid metals.
\newblock \emph{Phys. Rev. B}, 47:\penalty0 558--561, 1993.

\bibitem[Kresse and Hafner(1994{\natexlab{a}})]{kresse_ab_1994}
G.~Kresse and J.~Hafner.
\newblock \textit{{Ab} initio} molecular-dynamics simulation of the
  liquid-metal–amorphous-semiconductor transition in germanium.
\newblock \emph{Phys. Rev. B}, 49:\penalty0 14251--14269, 1994{\natexlab{a}}.

\bibitem[Kresse and Hafner(1994{\natexlab{b}})]{kresse_norm-conserving_1994}
G~Kresse and J~Hafner.
\newblock Norm-conserving and ultrasoft pseudopotentials for first-row and
  transition elements.
\newblock \emph{J. Phys.: Condens. Matter}, 6:\penalty0 8245--8257,
  1994{\natexlab{b}}.

\bibitem[Kresse and Furthmüller(1996{\natexlab{a}})]{kresse_efficiency_1996}
G.~Kresse and J.~Furthmüller.
\newblock Efficiency of ab-initio total energy calculations for metals and
  semiconductors using a plane-wave basis set.
\newblock \emph{Computational Materials Science}, 6:\penalty0 15--50,
  1996{\natexlab{a}}.

\bibitem[Kresse and Furthmüller(1996{\natexlab{b}})]{kresse_efficient_1996}
G.~Kresse and J.~Furthmüller.
\newblock Efficient iterative schemes for \textit{ab initio} total-energy
  calculations using a plane-wave basis set.
\newblock \emph{Phys. Rev. B}, 54:\penalty0 11169--11186, 1996{\natexlab{b}}.

\bibitem[Kresse and Joubert(1999)]{kresse_ultrasoft_1999}
G.~Kresse and D.~Joubert.
\newblock From ultrasoft pseudopotentials to the projector augmented-wave
  method.
\newblock \emph{Physical Review B}, 59:\penalty0 1758--1775, 1999.

\bibitem[Dudarev et~al.(1998)Dudarev, Botton, Savrasov, Humphreys, and
  Sutton]{dudarev_electron-energy-loss_1998}
S.~L. Dudarev, G.~A. Botton, S.~Y. Savrasov, C.~J. Humphreys, and A.~P. Sutton.
\newblock Electron-energy-loss spectra and the structural stability of nickel
  oxide: {An} {LSDA}+{U} study.
\newblock \emph{Phys. Rev. B}, 57:\penalty0 1505--1509, 1998.
\newblock \doi{10.1103/PhysRevB.57.1505}.
\newblock URL \url{https://link.aps.org/doi/10.1103/PhysRevB.57.1505}.

\bibitem[Ling et~al.(2000)Ling, Argyriou, Wu, and Neumeier]{Ling_2000}
Christopher~D. Ling, Dimitri~N. Argyriou, Guoqing Wu, and J.J. Neumeier.
\newblock Neutron diffraction study of \ce{La3Ni2O7}: Structural relationships
  among $n=1$, $2$, and $3$ phases \ce{La_{n+1}Ni_nO_{3n+1}}.
\newblock \emph{Journal of Solid State Chemistry}, 152:\penalty0 517--525,
  2000.
\newblock \doi{https://doi.org/10.1006/jssc.2000.8721}.
\newblock URL
  \url{https://www.sciencedirect.com/science/article/pii/S0022459600987218}.

\bibitem[Pizzi et~al.(2020)Pizzi, Vitale, Arita, Blügel, Freimuth, Géranton,
  Gibertini, Gresch, Johnson, Koretsune, Ibañez-Azpiroz, Lee, Lihm, Marchand,
  Marrazzo, Mokrousov, Mustafa, Nohara, Nomura, Paulatto, Poncé, Ponweiser,
  Qiao, Thöle, Tsirkin, Wierzbowska, Marzari, Vanderbilt, Souza, Mostofi, and
  Yates]{pizzi_wannier90_2020}
Giovanni Pizzi, Valerio Vitale, Ryotaro Arita, Stefan Blügel, Frank Freimuth,
  Guillaume Géranton, Marco Gibertini, Dominik Gresch, Charles Johnson,
  Takashi Koretsune, Julen Ibañez-Azpiroz, Hyungjun Lee, Jae-Mo Lihm, Daniel
  Marchand, Antimo Marrazzo, Yuriy Mokrousov, Jamal~I Mustafa, Yoshiro Nohara,
  Yusuke Nomura, Lorenzo Paulatto, Samuel Poncé, Thomas Ponweiser, Junfeng
  Qiao, Florian Thöle, Stepan~S Tsirkin, Małgorzata Wierzbowska, Nicola
  Marzari, David Vanderbilt, Ivo Souza, Arash~A Mostofi, and Jonathan~R Yates.
\newblock Wannier90 as a community code: new features and applications.
\newblock \emph{J. Phys.: Condens. Matter}, 32:\penalty0 165902, 2020.
\newblock \doi{10.1088/1361-648X/ab51ff}.
\newblock URL
  \url{https://iopscience.iop.org/article/10.1088/1361-648X/ab51ff}.

\end{thebibliography}

\end{document}



\title{Supplementary Material for 'Imaging stripe dynamics in trilayer nickelate \ce{La4Ni3O10}'}

\author{Uladzislau Mikhailau}
\affiliation{SUPA, School of Physics and Astronomy, University of St Andrews, North Haugh, St Andrews, KY16 9SS, United Kingdom}

\author{Luke C. Rhodes}
\affiliation{SUPA, School of Physics and Astronomy, University of St Andrews, North Haugh, St Andrews, KY16 9SS, United Kingdom}

\author{Siri A. Berge}
\affiliation{SUPA, School of Physics and Astronomy, University of St Andrews, North Haugh, St Andrews, KY16 9SS, United Kingdom}

\author{Matthias Hepting}
\affiliation{Max-Planck-Institute for Solid State Research, Heisenbergstr. 1, 70569 Stuttgart, Germany}

\author{Masahiko Isobe}
\affiliation{Max-Planck-Institute for Solid State Research, Heisenbergstr. 1, 70569 Stuttgart, Germany}

\author{Carolina de Almeida Marques}
\affiliation{SUPA, School of Physics and Astronomy, University of St Andrews, North Haugh, St Andrews, KY16 9SS, United Kingdom}

\author{Pascal Puphal}
\affiliation{Max-Planck-Institute for Solid State Research, Heisenbergstr. 1, 70569 Stuttgart, Germany}

\author{Peter Wahl}
\email[Correspondence to: ]{wahl@st-andrews.ac.uk.}
\affiliation{SUPA, School of Physics and Astronomy, University of St Andrews, North Haugh, St Andrews, KY16 9SS, United Kingdom}
\affiliation{Physikalisches Institut, Universität Bonn, Nussallee 12, 53115 Bonn, Germany}

\date{\today}

\maketitle

\section{Surface homogeneity}

\begin{figure}
\includegraphics[width=\textwidth]{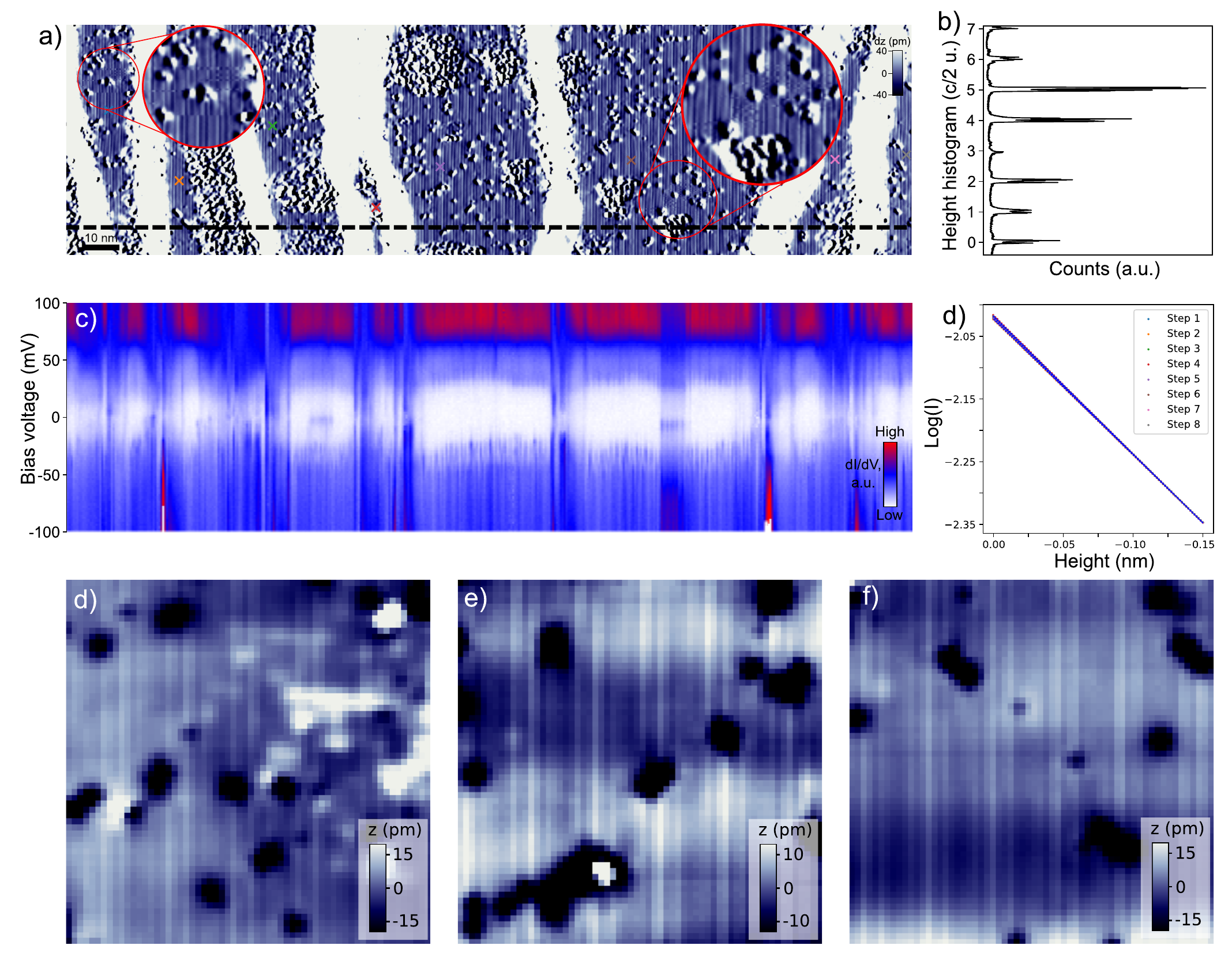}
    \caption{\textbf{Stripe order across terraces}. (a) Spatial derivative of a topography collected across multiple terraces ($V=200\mathrm{mV}$, $I=30\mathrm{pA}$). Areas of fluctuating stripes are highlighted by the red circles. Crosses show positions where $z(V)$-spectroscopy data was collected. (b) Histogram of the topography in (a), the step height is $\Delta z\sim 1.3\mathrm{nm}$. (c) Bias spectra $g(\mathbf{r},V)$ collected along the dashed line from (a) ($V=200\mathrm{mV}$, $I=50\mathrm{pA}$). (d) $\log(I(z))$ curves collected over multiple different terraces. (e-g) Zoomed in topographies extracted from the data shown in panel (a), highlighting commensurability of the modulation along the $c$ axis.}
    \label{SI:Steps}
\end{figure}

Only one surface termination has been observed in the STM images across multiple samples investigated. The step height, work function, and differential conductance are consistent between them (see Fig.~\ref{SI:Steps}). The average work function estimated from the slope of the $I(z)$ curves is $\Phi\simeq 4.7 \mathrm{eV}$. On every step edge stripes organize into commensurate domains with no incommensuration in the $c$ direction observed. 

\section{Fugacity and electron doping}
\label{fugacitydoping}
In reference \cite{anderson_theory_2006}, the asymmetry in the tunneling spectra is linked to the hole doping in a Mott-Hubbard model through
\begin{equation}
    Z(\mathbf{r},V)=\frac{g(\mathbf{r},V)}{g(\mathbf{r},-V)}=\frac{2x}{1-x}.
\end{equation}

This can, equivalently, be rewritten in terms of electron doping $n_\mathrm{el}$ through
\begin{equation}
    Z^\prime(\mathbf{r},V)=\frac{g(\mathbf{r},-V)}{g(\mathbf{r},V)}=\frac{2n_\mathrm{el}}{1-n_\mathrm{el}}.
\end{equation}
We can rewrite this in terms of the electron doping $n_\mathrm{el}$ through
\begin{equation}
    n_\mathrm{el}=\frac{Z^\prime(\mathbf{r},V)}{2-
    Z^\prime(\mathbf{r},V)}.
\end{equation}
This is the quantity plotted in figs.~2k and 3f.

\section{Phase locking of the stripe order}

\begin{figure}
\includegraphics[width=\textwidth]{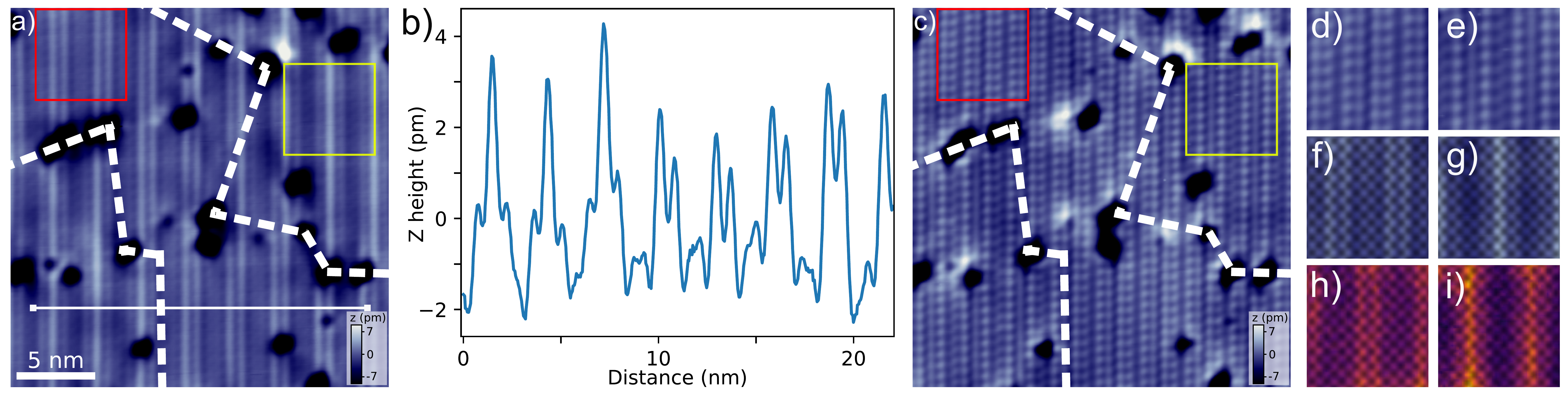}  \caption{\textbf{Phase-locking of stripes} (a) Large scale topography with setpoint conditions ($V=150\mathrm{mV}$, $I=40\mathrm{pA}$). Dashed line highlight boundaries in between different commensurately aligned regions. Red and yellow rectangles cover small defect-free areas with different alignment. (b) Linecut collected over the solid white line in image (a). (c) Topography in the same region with setpoint conditions ($V=-150\mathrm{mV}$, $I=40\mathrm{pA}$). (d-e) Zoomed-in rectangles from panel (c). (f-g) Zoomed in rectangles from panel (a),  with nickel lattice superimposed on top. (h-i) Calculated electron density $n_{el}(V=84~\mathrm{meV})$ with atomic positions superimposed on top. }
    \label{SI:Phase}
\end{figure}

As can be observed in Fig.~\ref{SI:Phase}(a), stripe order prefers to lock into a commensurate alignment with the lattice. Images are dominated by two different commensurate alignments: their structure is clearly seen in defect-free areas, marked by rectangles of different coloring Fig.~\ref{SI:Phase}(a). Fig.~\ref{SI:Phase}(a) depicts positive bias voltage height profile of two such regions with a dashed line indicating the boundary between them. Overall, such order can be characterised as locally commensurate, in the same way as stripes in cuprates. The size of the locally commensurate regions is spatially inhomogeneous and varies between samples. In Fig.~\ref{SI:Phase}(a) size of the locally commensurate regions is on the order of ten nanometers, while on cleaner patches present in Fig.~\ref{SI:Steps} it increases up to tens of nanometers.

Fig.~\ref{SI:Phase}(b) demonstrates that the modulation $\mathbf{q}_4=(1,1/4)$, which dominates at negative bias voltage also locks into locally commensurate alignments, similar to the stripes at positive bias. Zoomed in images of the alignments of the $\mathbf{q}_4=(1,1/4)$ modulation are shown in Fig.~\ref{SI:Phase}(d, e). Fig.~\ref{SI:Phase}(f, g) show commensurate patterns of charge modulation at the positive bias voltage in the same regions. By superimposing Ni atoms lattice modulation on top of the topography, we can observe that stripes always center between nickel atoms, where apical and in plane oxygens are situated. Fig.~\ref{SI:Phase}(h, i) present calculated electron doping for the same regions. Lines of increased electron doping are also centered on oxygen zigzags, out-of-phase to the charge modulation observed at positive bias voltage. 

\section{Calculations}
\label{calculations}

\begin{figure}
\includegraphics[width=0.6\textwidth]{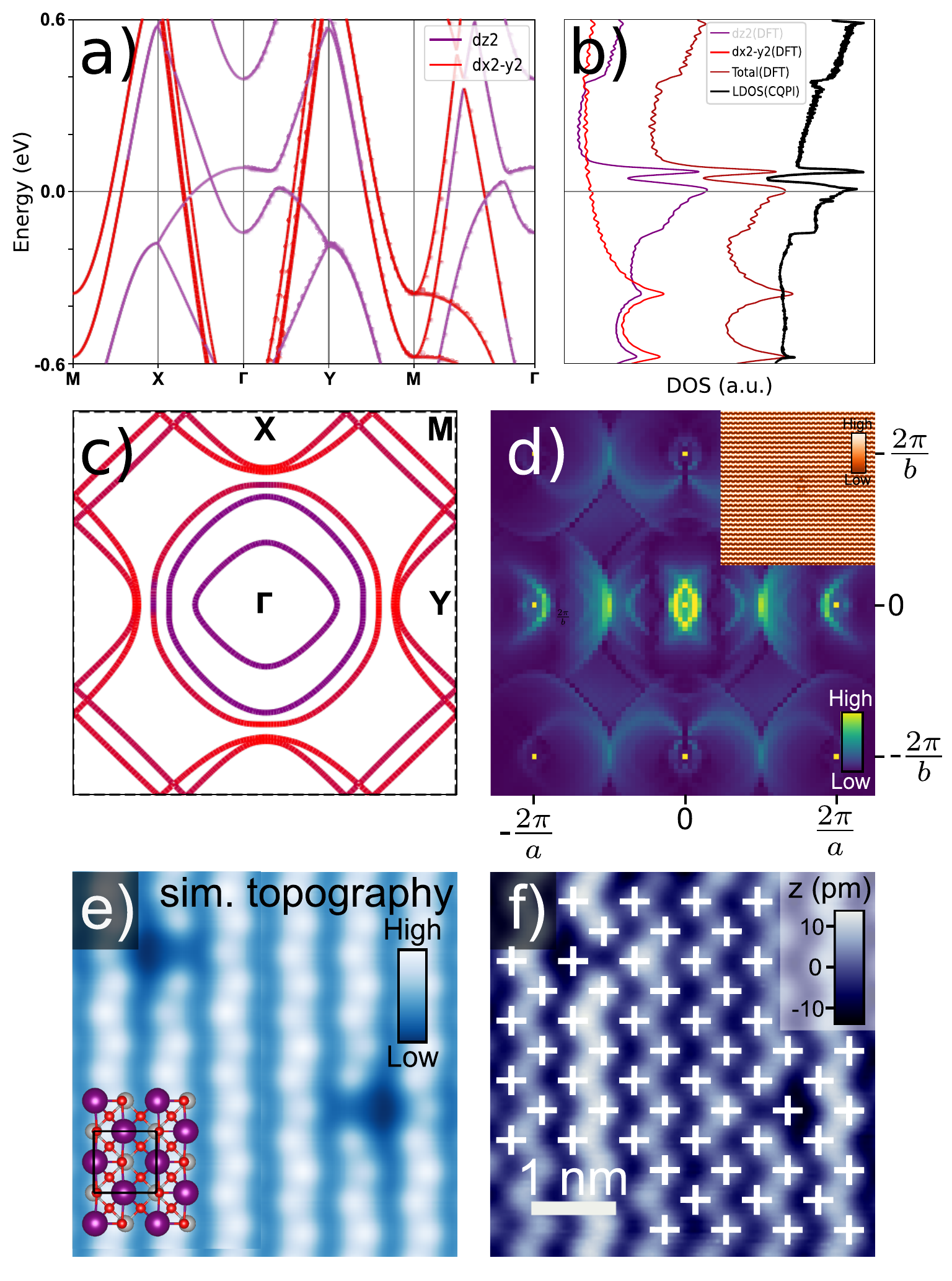}
    \caption{\textbf{DFT and cLDOS calculations} (a) Band structure of the monolayer of $\ce{La4Ni3O10}$ obtained from DFT+U calculations  (points), projected onto a tight-binding model using Wannier90 (lines). Color encodes orbital character. (b) Density of states calculated from the tight-binding model. Black line denotes local density of states calculated using calcQPI. (c) Fermi surface for the band structure as shown in (a). (d) Corresponding CQPI LDOS slice and its FFT image at the Fermi level. (e) Simulated topography with defects. (f) Experimental data, topography collected with setpoint conditions $V=150\mathrm{mV}$, $I=50\mathrm{pA}$. White crosses mark inferred positions of nickel atoms.}
    \label{SI:DFT}
\end{figure}

To determine the positions of the nickel atoms in STM topographies, we have performed continuum LDOS calculations using calcQPI\cite{wahl_calcqpi_2025}. For these calculations, we start from a Density Functional Theory (DFT) calculation of a slab using VASP. The calculations have been performed using the PBE functional \cite{kresse_ab_1993,kresse_ab_1994,kresse_norm-conserving_1994,kresse_efficiency_1996,kresse_efficient_1996,kresse_ultrasoft_1999} and with a plane wave energy cut off of $600\mathrm{eV}$, a $\mathbf{k}$-point grid of $7\times 7\times 1$, and including a Coulomb repulsion $U$ on the Ni $3d$ orbitals using $U-J=3\mathrm{eV}$ using the approach introduced by Dudarev\cite{dudarev_electron-energy-loss_1998}. The structure has been relaxed, starting from the crystal structure obtained from neutron diffraction.\cite{Ling_2000}
The DFT-calculated band structure is presented in fig.~\ref{SI:DFT}(a).
The calculation is then projected onto a tight-binding model with Wannier90\cite{pizzi_wannier90_2020}, which provides both the hopping terms as well as the localized wave functions. The resulting band structure, DOS and Fermi surface are plotted in Fig.~\ref{SI:DFT}(a-c). These are then fed into the QPI calculations to simulate the appearance of defects in STM images. Fig~\ref{SI:DFT}(e) shows such a simulated image, using an imaginary defect potential of $V=1.0~\mathrm{eV}$ to reproduce the appearance of the experimentally observed defects. The calculation readily reproduces the zig-zag modulation due to the octahedral tilts and shows good agreement with the experimental topography (Fig.~\ref{SI:DFT}(f)). Given the high similarity of experimental data and simulated picture, we can infer positions of the nickel atoms and highlight them in topographies as in fig.~\ref{SI:DFT}(f).

\section{Spin-polarised STM}
\begin{figure}
\includegraphics[width=\textwidth]{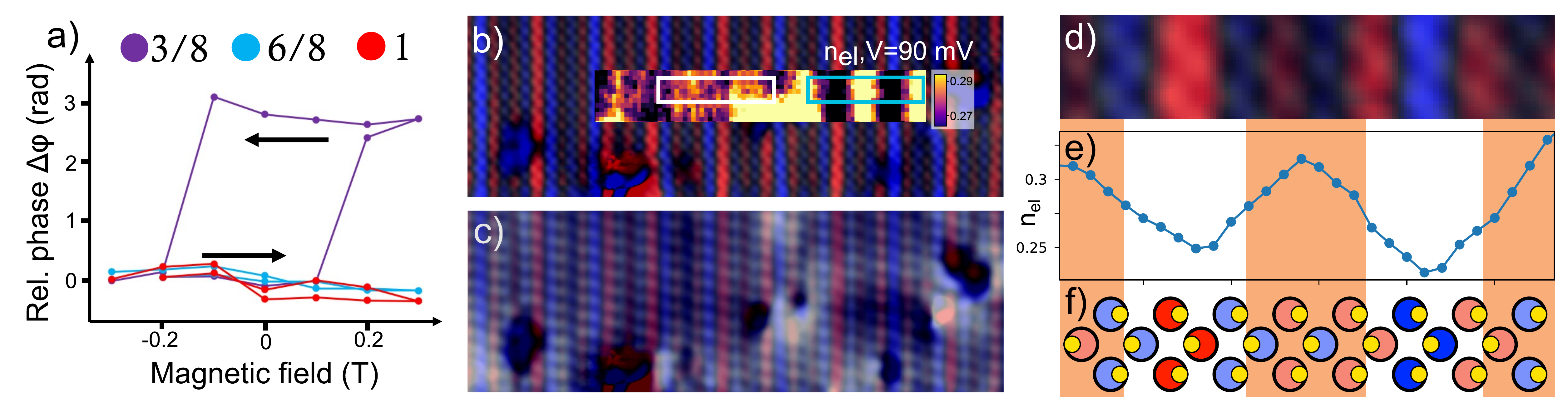}
    \caption{\textbf{Spin-polarised measurements} (a) Hysteresis loop of the tip based on the FFT analysis of topographies collected under applied magnetic field. It can be observed that structural and electronic modulations do not exhibit hysteretic behaviour, while magnetic modulation does. (b) Magnetic contrast with nickel lattice superimposed. The double stripe domain unit cell is shown in white, the single stripe domain unit cell is shown in cyan. Calculated $n_{el} (V=90\mathrm{mV})$ is shown in the inset. (c) Negative bias voltage topography, setpoint conditions are $V=-100\mathrm{mV}$, $I=20\mathrm{pA}$. (d) Magnetic modulation in the different domain with nickel lattice superimposed. (e) Row integrated electron doping $n_{el}(V=90\mathrm{mV})$. (f) Magnetic model with lines of diffuse electron doping.}
    \label{SI:SPIN}
\end{figure}

To characterize the magnetic properties of the tip of the STM, we have recorded a hysteresis loop of the ferromagnetic tip on the surface of $\ce{La4Ni3O10}$ (see Fig.~\ref{SI:SPIN}). Under application of out of plane magnetic field magnetization of the tip flips above $B=0.2~T$. This results in a global phase shift of the magnetic contrast at $\mathbf{q}_M = (0,3/8)$, while the electronic contrast at $\mathbf{q}_3 = (0,3/4)$ and structural features (e.g. at $\mathbf{q}_z = (0,1)$) are unaffected.

Fig.~\ref{SI:SPIN}(b) highlights commensurability of the magnetic order with nickel lattice. As shown in Fig.~\ref{SI:Phase}, the charge order is seen in regions of different commensurability. We find that where electronic domains change, the magnetic structure adjusts to a new commensurate pattern. As it can be inferred from Fig.~\ref{SI:Phase}(c), where magnetic contrast is superimposed over topography collected at negative bias voltage, bright doublets, coming from $\mathbf{q}_4=(1,1/4)$ modulation always link nickel atoms of different magnetic moment.

In Fig.~\ref{SI:SPIN}(d-f) we investigate the second domain structure. Compared to the order presented in the main text, it is a single high-intensity magnetic stripe arrangement with more diffused distribution of electron doping between of them.

\section{Stripe dynamics}

\begin{figure}
\includegraphics[width=0.6\textwidth]{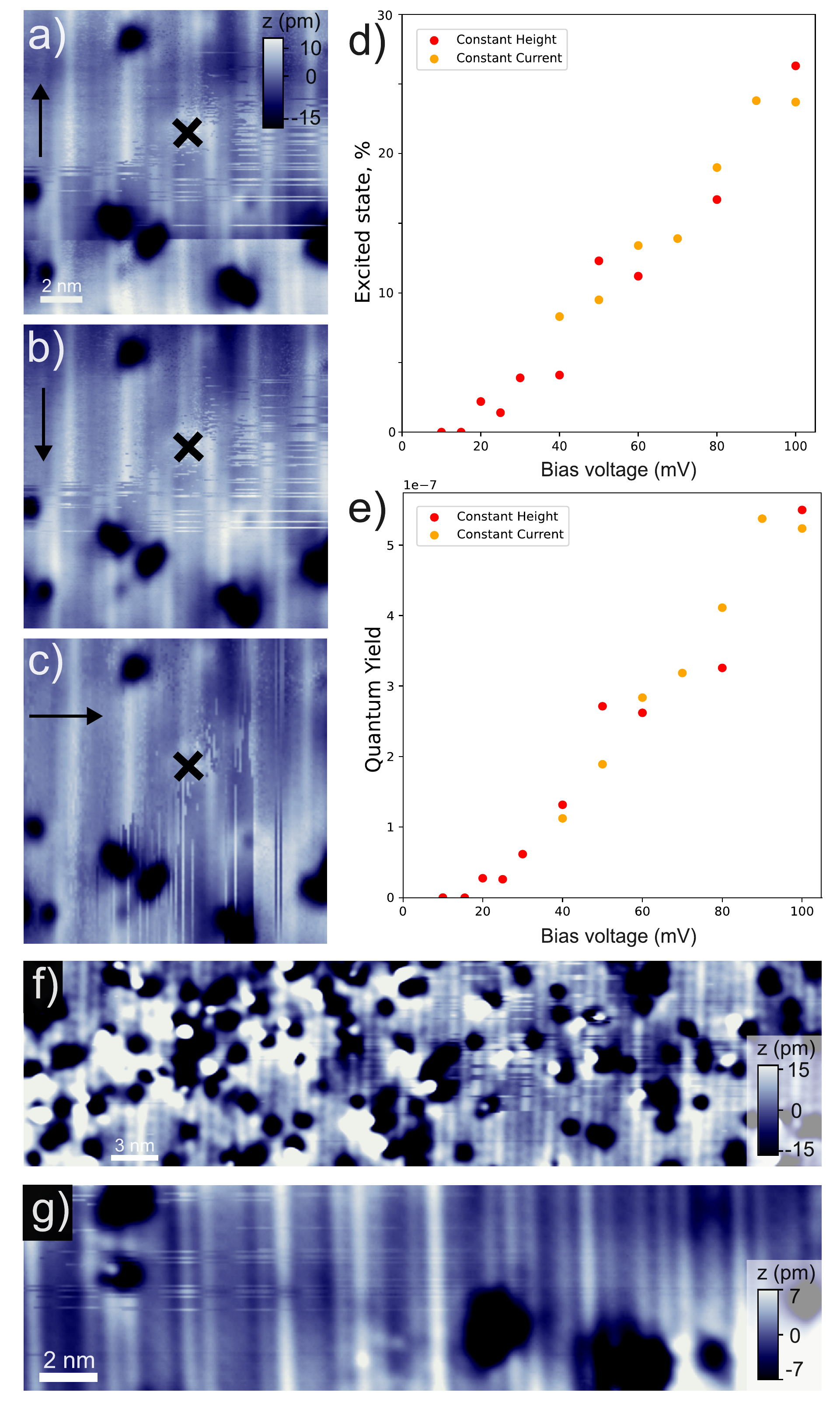}
    \caption{\textbf{Stripe Dynamics} (a-c) Collection of topographies at $V=100\mathrm{mV}$, $I=30\mathrm{pA}$ in the same place as in the main text. Scanning direction is highlighted by an arrow on the left side of the images. (d) Proportion the order is in an excited state in dependence to the applied bias voltage, measured with constant height and constant current. For constant height mode, the tip height is fixed at setpoint conditions $V_\mathrm{s} = 40~\mathrm{mV}$ and $I_\mathrm{s} = 30~\mathrm{pA}$ before ramping the bias voltage to the one where the time trace is recorded, for constant current measurements the tip is stabilized at $I_\mathrm{s} = 30~\mathrm{pA}$ at the bias voltage of the measurement before turning off the feedback loop and recording the time trace. (e) Quantum yield of the excitation depending on the applied bias voltage. Current traces for analysis were collected in the place depicted by cross on topographies. Parameters for constant height and current measurements as in (d). (f) Stripe dynamics imaged on a different crystal with a tip prepared on gold, setpoint conditions are $V=200\mathrm{mV}$, $I=20\mathrm{pA}$. (g) Stripe dynamics imaged with magnetic tip without external field, setpoint conditions are $150\mathrm{mV}$, $I=20\mathrm{pA}$.}
    \label{SI:Dynamics}
\end{figure}

Here we show additional data for the tip-induced stripe dynamics. We find that the fluctuations are insensitive to the scanning direction, see Fig.~\ref{SI:Dynamics}(a-c), i.e. the fluctuations do not simply follow the STM tip, but instead start to fluctuate. We call the ground state the state associated with the most frequent current in the histogram of a time trace (i.e. the mode), while the other states are considered excited states, which we only observe when the energy of the tunneling electrons exceeds the threshold of $\sim 15-20 \mathrm{meV}$. This can be seen in Fig.~\ref{SI:Dynamics}(d, e), where for voltages right above the threshold, both the proportion of time the system in the excited current states and the quantum yield of the process rise with increasing voltage. Quantum yield is the proportion of electrons tunneling from the tip which induce a transition to an excited state (i.e. $Y=\nu_\mathrm{exc}\cdot \frac{e}{I}$, where $\nu_\mathrm{exc}$ is the frequency of transitions to the excited states). The excited state proportion is the fraction of the time the systems is in one of the excited states. Both parameters rise linearly with increasing bias voltage, in line with the expectation for an effect driven by tunneling electrons. Notably, the effect does not depend on the distance between the tip and the surface, as both parameters do not change between constant height and constant current measurements. This practically excludes transient effects happening in the tunneling junction.

Additional data of stripe dynamics obtained on a different sample and with a different tip as well as with a ferromagnetic tip are shown in Fig.~\ref{SI:Dynamics}(f, g), respectively.

The time system spent in the high current state was obtained by using a threshold around the dominant amplitude above the noise level of the tunneling current. The frequency of fluctuations has been measured as a number of spikes above noise level in the time derivative of the current trace over a given time interval.

\label{Bibliography}
\bibliographystyle{unsrtnat}
\bibliography{la4ni3o10}